\documentclass[12pt]{article}
\usepackage{amssymb}
\usepackage{amsfonts}
\usepackage[active]{srcltx} 
\usepackage{amscd}

\newcommand{\adres}[1]{  {\small\it  \begin{center} #1 \end{center}   }     }
\newcommand{\dzieki}{ \begin{center}{\bf Acknowledgments} \end{center}  }

\newcommand{\dsp}[1]{\mbox{$(#1,{\cal #1})$}}

\newcommand{\fun}[3]{\mbox{$#1\colon #2 \- \longrightarrow #3$}}
\newcommand{\ww}{\dsp{W}}

\newenvironment{dow}{Proof.\hspace{0.5em}}{\hfill $\square$}
\newtheorem{tw}{Theorem}[section]
\newtheorem{lem}{Lemma}[section]
\newtheorem{stw}{Proposition}[section]
\newtheorem{defin}{Definition}[section]

\begin{document}




\title{\bf    Smooth Beginning of the  Universe}
\author{Jacek Gruszczak \thanks{email: sfgruszc@cyf-kr.edu.pl} }
\maketitle \adres{Institute of Physics,
       Pedagogical University, ul. Podchorazych 2,
       Cracow,Poland,
       and \\
       Copernicus Center for Interdisciplinary Studies, ul. Slawkowska 17, Cracow, Poland}


\begin{abstract}
The breaking down of the equivalence principle, when discussed in the
context of Sikorski's differential space theory, leads to the
definition of the so-called differentially singular boundary
(d-boundary) and to the concept of differential space with
singularity  associated with  a given space-time differential
manifold. This enables us to define the time orientability, the
beginning of the cosmological time and the
 smooth evolution for the flat
FRW world model with the initial singularity. The simplest smoothly
evolved models are studied. It is shown, that the cosmological matter
causing  such an evolution can be of three different types. One of
them is the fluid with dark energy properties, the second the fluid
with attraction properties, and the third a mixture of the other two.
Among  all investigated smoothly evolved solutions, models
qualitatively consistent with the observational data of type Ia
supernovae have been found.
\end{abstract}



\maketitle


\section{Introduction}
\label{wstep} It is generally believed that the Universe had a
beginning and everything indicates   that it indeed had the beginning.
According to  contemporary ideas the Planck era is the beginning of the
Universe. During the flow of cosmic time the known Universe emerges
from the Planck era. Certainly it is  not an immediate process and one
can imagine that individual space-time structures emerge from the
Planck era gradually. One can expect that the simplest structures such
as a set structure or a topology on this set can emerge first. The
appearance of the manifold structure is the key event of the process.
Then the equivalence principle, in its non metric form, appears and
makes possible a creation of '' higher'' structures such as, for
example, the Lorentzian structure. From this perspective it is
interesting to ask how the space-time geometry is ''nested'' in
theories more general than the differential geometry since the
emerging process of gravitation has to be associated with a theory
which is more general than the geometry of space-time manifold.

In this paper the Sikorski's differential spaces theory (see Appendix A) is applied
to the discussion of the breaking down process of the equivalence principle. Within
this theory the non-metric version of the equivalence principle is one of the axioms
of the space-time differential manifold (d-manifold) definition. Let us consider
points at which this axiom is not satisfied. The set of this points forms the
so-called differentially singular boundary (d-boundary). This issue is discussed in
Section 2. The construction of this type of boundary is also described.

In Section 3, the construction from Section 2 is used for building the
differential space (d-space) with d-boundary  for the flat FRW model.
The issue of the prolongation of cosmic time and time-orientability to
the d-boundary for this model is discussed in Section 4. In Section 5,
we describe the concept of the smooth evolution of this model starting
from the initial singularity. We find the simplest smoothly evolved
flat FRW models in Section 6. Some of them (see Section \ref{dark})
have  the kinematical properties which are qualitatively agreeing with
the recent cosmological observations \cite{RiessAcceleration,
PerlAcceleration}.  Namely, these models begin their evolution with
the decelerate expansion which changes into the accelerated one.

The discussion of properties of the cosmological fluid which causes that type of
smooth evolution  for the simplest cosmological models is carried out in Section 7.
Behaviour of any smoothly evolving flat model in a neighbourhood of the initial
singularity is investigated in Section 8. Finally, in Section 9, we summarize  the
main results of this paper


\section{Sikorski's differential spaces and singularities}
\label{dsing} Space-time is a 4-dimensional, Lorentzian and time-orientable
d-manifold of the class  $C^\infty$  \cite{9bo}. This definition is a mathematical
synthesis of what is known about  gravitational field. It implicitly includes the
equivalence principle  \cite{15bo,16bo,17bo}.

The equivalence principle, in its non-metric version, is implicite in the axiom
stating that space-time is a d-manifold which means that it is  locally homeomorphic
to an open set in $\mathbb{R}^n$, where $n=4$. Differential properties of these
homeomorphisms (maps) are determined by the atlas axioms \cite{kobayashi} through the
assumption, that the composition of two maps $\varphi_1\circ\varphi_2^{-1}$, as a
mapping between open subsets of $\mathbb{R}^n$, is a diffeomorphism of the class
$C^k$, $k\in \mathbb{N}$. In the present paper it is assumed that these
diffeomorphisms are smooth. It is worth to notice, that the non-metric version of the
equivalence principle is encoded into two levels of the classical d-manifold
definition, namely in  the assumption of the existence of local homeomorphisms to
$\mathbb{R}^n$ and  in the axioms of atlases.

One can look at  space-time, or more generally at a d-manifold,   from
the viewpoint of theories that are more general than  classical
differential geometry. In these theories d-manifolds are special
examples of more general objects. Examples of theories of this type
are: the theory of sub-cartesian spaces   by Aronszajn and Marshall
\cite{aronszajn1,27am}, the theory of Mostow's spaces \cite{28am} or
Chen's theory \cite{61am}. In the present paper we  study the initial
singularity problem and the problem of the beginning of cosmological
time  with help of Sikorski's differential spaces (d-spaces for
brevity) \cite{3bo, 34af, 35af}. In this theory, the generalization
level does not lead to an excessive abstraction and therefore the
d-space's theory may have applications in physics.

Generally speaking,  d-space is a pair  $(M,{\cal M})$, where $M$ is any set, and
${\cal M}$ is a family of functions $\fun{\varphi}{M}{\mathbb{R}} $ (see Appendix
\ref{appendix}). This family satisfies the following conditions: a) ${\cal M}$ is
closed with respect to superposition with smooth functions from  $C^\infty
(\mathbb{R}^n,\mathbb{R})$, and b) ${\cal M}$ is closed with respect to localization.
The precise sense of these notions is not important at the moment. The  d-space
$(M,{\cal M})$ is also a topological space $(M,\tau_{\cal M})$, where  $\tau_{\cal
M}$ denotes the induced topology on $M$ given by the family ${\cal M}$. The topology
$\tau_{\cal M}$ is the weakest topology in which all functions from ${\cal M}$ are
continuous. With in the theory of d-spaces the d-manifold definition assumes the form

\begin{defin}
\label{dmanifold} Non empty d-space $(M, {\cal M})$ is n-dimensional d-manifold, if
the following condition is satisfied
\begin{itemize}
\item[{\rm(}$\star${\rm )}] for every $p\in M$, there are neighbourhoods  $U_p\in \tau_{\cal M}$ and $O\in
\tau_{\mathbb{R}^n}$, such that d-subspaces $(U_p,{\cal M}_{U_p})$ and $(O,{\cal
E}^{(n)}_O)$ of d-spaces $(M, {\cal M})$ and $(\mathbb{R}^n,{\cal E}^{(n)})$
respectively
 are diffeomorphic, where ${\cal E}^{(n)}:=C^\infty
(\mathbb{R}^n,\mathbb{R})$.
\end{itemize}
\end{defin}
Briefly speaking,  d-space $(M, {\cal M})$ is a n-dimensional
d-manifold if it is locally diffeomorphic to  open subspaces of
$\mathbb{R}^n$ which are treated as d-spaces (see Definitions
\ref{locdyf} and \ref{dman}). Diffeomorphisms appearing in condition
($\star$) are generelized maps. Unlike in the classical definition,
these maps  are automatically diffeomorphisms. Definition
\ref{dmanifold} is equivalent to the classical d-manifold definition
\cite{3bo}.

In the d-spaces theory,  the non-metric version of the equivalence
principle is ''localized'' in the single axiom ($\star$). This enables
us to give the following definition of space-time

\begin{defin} \label{newdef} A space-time is a non empty d-space  $(M, {\cal M})$ such that
\begin{itemize}
\item[a)] $(M, {\cal M})$ satisfies the  non-metric equivalence principle as expressed by the condition ($\star$)
for n=4,
\item[b)] a Lorentzian metric form $g$ is defined on $M$,
\item[c)] Lorentzian d-manifold  $(M, {\cal M}, g)$ is time orientable.
\end{itemize}
\end{defin}

If   the non-metric equivalence principle in the classical definition
of space-time is thrown out, the time-orientability, the Lorentzian
metric structure, the classical differential structure and the
topological manifold structure are    automatically destroyed.  Only
the topological space structure survives.

In the case  space-time Definition \ref{newdef} the situation is
different. Throwing out  the non-metric equivalence principle  in the
form of  axiom ($\star$) leaves a rich and workable structure on which
one can  define generalized counterparts of classical notions such as
orientability, Riemannian and pseudo-Riemannian metrics, tensors, etc
\cite{3bo, 34af, 35af, 12af,31am,rosinger2,megied1, buchner1, heller1}.

Looking at space-time manifolds from the d-spaces perspective, one can imagine a
situation when a background of our considerations is a sufficiently ''broad'' d-space
$(M, {\cal M})$ which is not a 4-dimensional d-manifold. Let us additionally suppose
 that in $(M, {\cal M})$ there exists a subset  $B\subset M$ having a structure of
   a 4-dimensional d-manifold satisfying the condition ($\star$) for
$n=4$, regarded as a d-subspace of $(M,{\cal M})$. Naturally, at each point $p\in B$
the non-metric equivalence principle is satisfied.  The points not belonging to $B$
 can considered as singular points because at these points the four-dimensional
non-metric equivalence principle is violated. Among such singular points in $M$ the
most interesting are accumulation points of the set $B$ in the topological space $(M,
\tau_{\cal M})$. One can call a set of all singular accumulation points of $B$ the
singular boundary of  $(B,{\cal M}_{B})$ space-time d-manifold. Singular points in
$M$ not belonging to the singular boundary  can be excluded from considerations as
unattainable from the interior $B$ of the space-time d-manifold.

\begin{defin}\label{dbrzeg}
Let $(M,{\cal M})$ and $(M_0,{\cal M}_0)$ be Sikorski's d-spaces, such that
\begin{itemize}
\item[a)] $(M_0,{\cal M}_0)$ is a n-dimensional d-manifold,
\item[b)] $(M,{\cal M})$ is not a   n-dimensional d-manifold,
\item[c)] $M_0\subset M$ and $M_0$ is a dense set in the topological space
 $(M,\tau_M)$,
\item[d)] $(M_0,{\cal M}_0)$ is a d-subspace of the d-space
 $(M,{\cal M})$.
 \end{itemize}
The set $\partial_d M_0:=M-M_0$ is said to be differentially-singular boundary
(d-boundary for brevity) of d-manifold $(M_0,{\cal
 M}_0)$,
 if for every $p\in \partial_d M_0$ and for every neighbourhood $U_p\in \tau_{\cal
 M}$, the d-subspace $(U_p,{\cal M}_{U_p})$ is not diffeomorphic to
 any d-subspace  $(V,C^\infty({\mathbb R}^n)_V)$ of the d-space
  $(\mathbb{R}^n, C^\infty(\mathbb{R}^n))$, where $V\in
 \tau_{\mathbb{R}^n}$.
 Then, the d-space $(M,{\cal M})$ is said to be  d-space with differentially singular boundary
   (or d-space with d-boundary for brevity) associated with the d-manifold  $(M_0,{\cal M}_0)$.
\end{defin}

In the following we shall restrict our considerations to the case of
Definition \ref{dbrzeg} for $n=4$. The d-boundary definition for a
space-time's d-space $(M_0,{\cal M}_0)$ depends on the choice of the
d-space $(M,{\cal M})$.  Therefore, a reasonable method of the d-space
$(M,{\cal M})$ construction is necessary. Fortunately, the idea
described above  suggests such a construction. Namely, one chooses a
sufficiently ''broad'' d-space ''well'' surrounding the whole
investigated space-time $M_0$ and  then one carries out the process of
determination of all accumulation points for $M_0$. These points form
the  d-boundary $\partial_d M_0$ of space-time. Next, one defines  the
set $M$  as $M:=M_0\cup\partial_d M_0$. By treating $M$ as a
d-subspace of the sufficiently ''broad'' d-space, the d-space with the
d-boundary $(M,{\cal M})$ is uniquely determined. On can easily check
that the pair of d-spaces $(M,{\cal M})$ and $(M_0,{\cal M}_0)$, where
$(M,{\cal M})$ is determined by above described method, satisfies
Definition \ref{dbrzeg}.

The choice of a sufficiently ''broad'' d-space ''well'' surrounded the whole
investigated space-time is almost obvious. As it is well known \cite{3af},  every
space-time can be globally and isometrically embedded in a sufficiently dimensional
pseudo-Euclidean space $E^{p,q}$, where $p,q$ depends on space-time model. The space
$E^{p,q}$ is also  the  d-space $(\mathbb{R}^{p+q},C^\infty
(\mathbb{R}^{p+q},\mathbb{R}))$ which ''well'' surrounds the studied space-time
d-manifold $(M_0,{\cal M}_0)$.

Such a choice of the surrounding d-space is well motivated since all causal properties
of the studied space-time are taken into account, even if we do not refer to them
directly. This is because of the isometricity  of the embedding. Furthermore, the
concrete form of the isometrical embedding provides us with a practical method of
constructing the d-structure ${\cal M}$ for the the d-space $(M,{\cal M})$. Namely,
${\cal M}=C^\infty (\mathbb{R}^{p+q},\mathbb{R})_M$.

To test this method of constructing $(M,{\cal M})$ we shall apply it to   the flat FRW
cosmological model.


\section{A differential space for the flat FRW model with singularity}
\label{dsp}

Let us consider the flat FRW model with the metric
\begin{equation}\label{metryka}
g=-dt^2+a_{0}^2(t)(dx^2+dy^2+dz^2),
\end{equation}
 where $(t,x,y,z)\in
 W^{0}:=D_a^{0}\times\mathbb{R}^3$ and $D_a^{0}$ is a domain of the scale factor .
  The scale factor \fun{a_{0}}{D_a^{0}\ni
 t}{a_{0}(t)\in\mathbb{R}^+}, $a_{0}\in C^\infty(D_a^{0})$,
is an even real function of the cosmological time  $t$. A domain of the scale factor
 $D_a^{0} \subset\mathbb{R}^+,$   is an open and connected set. For convenience,
 let us assume that $t=0$
is an accumulation point of the set $D_a^{0}$ and  $0\notin D_a^{0}$.  In
addition, the model has an initial singularity at $t=0$ i.e. $\lim_{t\rightarrow
0^+}a_{0}(t)=0$. In the next parts of the paper, the symbol $a$ is reserved for
the following function: $\fun{a}{D_a}{\mathbb{R}}$, $a(t)=a_{0}(t)$ for $t\in
D_a^{0}$, $a(t)=0$ for $t=0$ where $D_a:=D_a^{0}\cup \{ 0\}$.

A construction of a d-space with d-boundary for the flat FRW model  pass in
analogical way to the construction of a differential space for the string-generated
space time with a conical singularity described in \cite{69af}.

Every 4-dimensional Lorentzian manifold can be  isometrically embedded in a
pseudo-Euclidean space \cite{3af}. In particular, the manifold $(W^{0},g)$ of
model (\ref{metryka}) can be isometrically embedded in $(\mathbb{R}^5,\eta^{(5)})$
 by means of the  following mapping
$$\fun{F^{0}}{W^{0}}{F^{0}(W^{0})}\subset \mathbb{R}^5$$
\begin{equation}
\begin{array}{c}
F^{0}_1(t,x,y,z)=\frac{1}{2}a_{0}(t)(x^2+y^2+z^2+1)+\frac{1}{2}\int_{0}^{t}\frac{d\tau}{\dot{a}_{0}(\tau)},
\\
F^{0}_2(t,x,y,z)=\frac{1}{2}a_{0}(t)(x^2+y^2+z^2-1)+\frac{1}{2}\int_{0}^{t}\frac{d\tau}{\dot{a}_{0}(\tau)},\\
F^{0}_3(t,x,y,z)=a_{0}(t)x, \hspace{1em}
F^{0}_4(t,x,y,z)=a_{0}(t)y,\hspace{1em} F^{0}_5(t,x,y,z)=a_{0}(t)z,
\end{array} \label{zanurzenie}
\end{equation}
where $\eta^{(5)}=diag(-1, 1, 1, 1, 1)$.

As every manifold,  space-time $(W^{0},g)$ is also a differential space
 $(W^{0}, {\cal W}^{0})$, where the differential structure
${\cal W}^{0}:={\cal E}^{(4)}_{W^{0}}$ is a family of local ${\cal
E}^{(4)}$-functions on  $W^{0}\subset\mathbb{R}^4$, where ${\cal
E}^{(4)}=C^\infty(\mathbb{R}^4)$. On the other hand, the set
$F^{0}(W^{0})$ can be equipped, in a natural way, with a  differential
structure treated
  as a differential subspace of the $(\mathbb{R}^5, {\cal E}
^{(5)})$, where ${\cal E}^{(5)}=C^\infty(\mathbb{R}^5)$. Then the family ${\cal E}
^{(5)}_{F^{0}(W^{0})}$ of local ${\cal E} ^{(5)}$-functions on ${F^{0}(W^{0})}$  is a
differential structure and the pair $({F^{0}(W^{0})},{\cal E}
^{(5)}_{F^{0}(W^{0})})$  is a differential space.  In addition, if the integral
$\int_0^t d\tau /\dot{a}_0(\tau)$ is convergent for every $t\in D_0$, then the
mapping $F^0$ is a diffeomorphism of the differential spaces $(F^0 (W^0),{\cal E
}^{(5)}_{F^0 (W^0)})$ and $(W^0 , {\cal W}^0)$ \cite{30am,69af}.

The process of attaching the initial singularity   depends on  the completion of the
set $F^0(W^0)$ by means of points from the surrounding space $\mathbb{R}^5$ in the
way controlled by the isometry $F^0$. It enables us to define the d-space with
d-boundary (with the initial singularity) for the flat FRW model.

Let the mapping \fun{F}{W}{\mathbb{R}^5} denotes a prolongation of
 $F^{0}$  to the set $W:=D_a\times\mathbb{R}^3$. The values of $F$ are given
 by  formulae
 (\ref{zanurzenie}) changing the symbol $F^{0}$ onto $F$.
 Then for every $x,y,z\in \mathbb{R}$ the value of $F$ at the initial moment $t=0$ is
 $F(0,x,y,z)=(0,0,0,0,0)$, since $a(0)=0$ by assumption. Therefore,  the cosmological initial
 singularity (d-boundary)
  distinguished by the embedding procedure
 is represented by the single point: $\partial_d F^0(W^0)=\{(0,0,0,0,0)\}\subset \mathbb{R}^5$.

The pseudo-Euclidean space $(\mathbb{R}^5, \eta^{(5)})$ is a differential space
$(\mathbb{R}^5, {\cal E}^{(5)})$. The differential structure ${\cal E}^{(5)}$ of this
space is finitely generated.  The projections on the axes of the Cartesian system
\fun{\pi_i}{\mathbb{R}^5}{\mathbb{R}}, $\pi_i(z_1,z_2,...,z_5)=z_i$, $i=1,2,...,5$,
are generators of this structure and  therefore: ${\cal E}^{(5)}={\rm Gen}(\pi_1 ,
\pi_2,...,\pi_5)$. As is well known \cite{12af,69af}, every subset $A$ of a support
$M$ of a differential space $(M, \cal{C})$ is a differential subspace  with the
following differential structure: ${\rm Gen}({\cal C}|_A)$. For finitely generated
differential spaces $(M, \cal{C})$, every differential subspace with a support
$A\subset M$ is also finitely generated.  Generators of the differential structure on
$A$ are generators of the differential structure ${\cal C}$ restricted to $A$.

The differential space $(F(W),{\cal E}^{(5)}_{F(W)})$ is a differential subspace of
the differential space $(\mathbb{R}^5, {\cal E}^{(5)})$ and represents the d-space
with d-boundary for model  (\ref{metryka}). Since ${\cal E}^{(5)}$ is finitely
generated, the d-structure ${\cal E}^{(5)}_{F(W)}$, induced on $F(W)$, is also
finitely generated and

$${\cal E}^{(5)}_{F(W)}={\rm
Gen}(\pi_1|_{F(W)},\pi_2|_{F(W)},...,\pi_5|_{F(W)}).$$


The d-space $(F(W),{\cal E}^{(5)}_{F(W)})$ is not convenient for further discussion
because it depends on the embedding procedure. Let us define a d-space
$(\bar{W},\bar{\cal W})$ diffeomorphic to $(F(W),{\cal E}^{(5)}_{F(W)})$ which  will
enable us to apply the Sikorski`s geometry in a form similar to the standard
differential geometry.

Let us consider an auxiliary d-space $(W,{\cal W})$,  ${\cal W}={\rm Gen}({\beta}_1,
{\beta}_2,...,{\beta}_5)$, where
$$\fun{{\beta}_i}{W}{\mathbb{R}},\hspace{1em}
{\beta}_i(t,x,y,z)=\pi_i\circ F(t,x,y,z), \hspace{1em} i=1,2,...,5.$$ This space is
not diffeomorphic to the d-space with d-boundary $(F(W),{\cal E}^{(5)}_{F(W)})$, since
the function \fun{F}{W}{\mathbb{R}^5} is not one to one. Additionally, the generators
${\beta}_i$ do not distinguish the following points $p\in\partial W:=W - W^0 $.
Therefore, the d-space $(W,{\cal W})$ is not a Hausdorff space. With the help of this
space one can build a d-space with d-boundary $(\bar{W},\bar{\cal W})$ diffeomorphic
to $(F(W),{\cal E}^{(5)}_{F(W)})$.

Let $\varrho_H$ be the following equivalence relation
$$\forall p, q:\hspace{1em} p \varrho_H  q \Leftrightarrow
\forall {\beta}\in{\cal W}: {\beta}(p)={\beta}(q). $$ The quotient space
$\bar{W}=W/\rho_H$ can be equipped with a d-structure $\bar{\cal W}:={\cal
  W}/\rho_H$ coinduced from {\cal W}
$$\bar{\cal W}:={\rm Gen}(\bar{\beta}_1, \bar{\beta}_2,...,
\bar{\beta}_5),$$ where
$$\fun{\bar{\beta}_i}{\bar{W}}{\mathbb{R}},\hspace{1em}
\bar{\beta}_i([p]):={\beta}_i(p),\hspace{1em} i=1,2,...,5.$$ The symbol $[p]$ denotes
the equivalence class of a point $p\in W$ with respect to $\varrho_H$. The pair
$(\bar{W}, \bar{\cal W})$ is a d-space \cite{58am,30am,69af}.

Let us define the following mapping
$$\fun{\bar{F}}{\bar{W}}{F(W)},\hspace{1em}  \bar{F}([p]):=F(p).$$

\begin{tw}\label{diffe}
The d-space $(\bar{W}, \bar{\cal W})$ is diffeomorphic to the d-space with
d-boundary   $(F(W),{\cal E}^{(5)}_{F(W)})$.\hfill $\blacksquare$
\end{tw}

\begin{dow}
 The mapping $\fun{\bar{F}}{\bar{W}}{F(W)}$ is a bijection. In addition,
 $\bar{W}$ is by  construction a Hausdorff topological space with the topology
given by the generators  $\bar{\beta}_1, \bar{\beta}_2,...,
\bar{\beta}_5$ \cite{3bo,34af,35af 12af}. The mapping
$$\bar{F}([p])=({\beta}_1(p), {\beta}_2(p),..., {\beta}_5(p))=
(\bar{\beta}_1([p]), \bar{\beta}_2([p]),..., \bar{\beta}_5([p]))$$
is a diffeomorphism of
the d-space $(\bar{W},\bar{\cal W})$ onto its image $(\bar{F}(\bar{W}),{\cal
E}^{(5)}_{\bar{F}(\bar{W})})=(F(W),{\cal E}^{(5)}_{F(W)})$ \cite{32af}.
\end{dow}

 According to Definition \ref{dbrzeg},  $(\bar{W},\bar{\cal W})$   is the d-space
with d-boundary for the flat FRW d-manifold  $(W^{0}, {\cal W}^{0})$, where the
d-boundary is represented by the set $\partial_d W^0:=\bar{W}-W^0$.

The d-space with d-boundary $(\bar{W},\bar{\cal W})$ has been constructed with help of
the d-space $(W,{\cal W})$ and the relation $\varrho_H$. Generally speaking, d-spaces
of the type of $(M,{\cal C})$ and  $(M/\varrho_H,{\cal C}/\varrho_H)$ have a lot of
common features because of the isomorphism of the algebras ${\cal C}$ i ${\cal
C}/\varrho_H$. In particular,  moduli of smooth vector fields $\textbf{X}(M) $ and
$\textbf{X}(M/\varrho_H)$ are isomorphic \cite{30am,69af}. This property will, in the
next parts of the paper, enable us to work   with the help of the more convenient
d-space $(W,{\cal W})$ instead of $(\bar{W},\bar{\cal W})$.

The d-structure ${\cal W}$ of the d-space  $(W,{\cal W})$ is finitely generated by
means of functions $\beta_i$, $i=1,2,...,5$. However, in the next parts of the paper,
we use  a different, but  equivalent, system of generators
\begin{equation}\label{generatory}
\alpha_1:=\beta_1-\beta_2,\hspace{1em} \alpha_2:=\beta_1+\beta_2,\hspace{1em}
\alpha_i:=\beta_i, \hspace{1em} i=3,4,5.
\end{equation}
Then the d-structure ${\cal W}$ has the form
$${\cal W}={\rm Gen}(\alpha_1, \alpha_2,...,\alpha_5).$$


\section{Time orientability  }
\label{wekt}

The flat FRW model  is a time orientable Lorentzian manifold $M$.  By definition,
there is a timelike directional field generated by a nowhere vanishing timelike
vector field $X$.   If $X$ generates the directional field then the field $\lambda X$
generates it also, where $\lambda$ is a nowhere vanishing scalar field on $M$. The
field $X$ caries a part of information included in the casual structure of $M$,
which  enables us  to define the  direction of the stream of time and the succession
of events \cite{70af}.

The manifold structure and the casual structures of  space-time  are broken down at
the initial singularity. In the hierarchy of space-time structures the Sikorski's
d-structure is placed
 below  the casual structure \cite{12af}. Therefore,
the d-space with d-boundary $(\bar{W},\bar{\cal W})$ of the flat FRW model is timeless
independently of the fact that  one of coordinates is called time and the moment
$t=0$ is named the beginning of time. In this situation one cannot say that the
cosmological singularity  (d-boundary) is an initial or  final state of the cosmic
evolution. It is necessary to introduce a notion which would be  a substitute of time
orientability.

Let $(W^{0}, {\cal W}^{0})$ and $(W, {\cal W})$ be the pair of d-spaces described in
the section \ref{dsp}.
 For  convenience we can consider the d-space $(W,{\cal W})$  instead of the
d-space with d-boundary $(\bar{W},\bar{\cal W})$ according to the remark after
Theorem \ref{diffe}. In this representation the set of not Hausdorff separated points
$\partial W:=W-W^0$ is a counterpart of the d-boundary $\partial_d W^0$ for the flat
FRW d-manifold.

Let in addition,
  $\fun{X^{0}}{W^{0}}{TW^{0}}$ be a timelike and smooth vector field
without critical points, tangent to the manifold
 $(W^{0}, {\cal W}^{0})$,   fixing the time-orientability on the manifold
 $(W^{0}, g)$.

\begin{defin}\label{czorie2}
The d-space $(W, {\cal W})$ is said to be  time  oriented  by means of a vector field
$X$ if
\begin{itemize}
\item[a)] there is a nonzero vector field $\fun{X}{W}{TW}$ tangent to $(W, {\cal W})$
given by the formula
$$\forall \alpha\in {\cal W}: \hspace{0.3em} X(p)(\alpha):=\left\{
\begin{array}{ccl}
X^0(p)(\alpha|_{W^0}) &{\it for}&p\in W^0 \\
\lim_{q\rightarrow p} X^{0}(q)(\alpha|_{W^{0}}) &{\it for}&  p\in \partial
W,\hspace{1em} q\in W^{0}
\end{array}\right.
$$
\item[b)]and there is a function $\lambda \in {\cal W}$,  $\lambda (q)>0$ (or $\lambda
(q)<0$) for $q\in W^0\subset W$ such that the vector field $V:=\lambda X$ is
smooth  on $(W, {\cal W})$.
\end{itemize}

A coordinate defined by means of  $X$ is called  time and the moment $t=0$ the
beginning of  time $t$. We also say that the d-space $(W, {\cal W})$ is oriented
with respect to  time $t$.
\end{defin}

In the flat FRW model (\ref{metryka}) the cosmological time  $t$ is a time variable.
The vector field of the form
\begin{equation}\label{poleX}
\fun{X^0}{W^{0}}{TW^{0}}, \hspace{1em} X^0
(p)(\alpha^0):=\frac{\partial \alpha^0 (p)}{\partial t},
\end{equation}
 where
$\alpha^0\in {\cal W}^{0}, p\in W^0$,  establishes the time orientation on $(W^0 ,
g)$. The vector field is  smooth  on the manifold $(W^0, {\cal W}^0 )$ since
derivation  $\fun{\widehat{X}^0}{{\cal W}^0}{\mathbb{R}^{W^0}}$, $\widehat{X}^0
(\alpha^0)(p):=X^0 (p)(\alpha^0)$     satisfies the  condition $\widehat{X}^0 ({\cal
W}^0 )\subset {\cal W}^0 $ (Definition \ref{polewektorowegladkie}).

 In the next parts of the  paper,  cosmological models  for which
 the vector field   $X^0$ can be extended on $(W, {\cal W})$ are discussed.
 Then if remaining conditions of  Definition \ref{czorie2} are satisfied,
    $\ww$ is a d-space with  boundary $\partial W$
    of the flat FRW d-manifold which is
   time oriented
   with help of the following vector field
\begin{equation}\label{poleY}
\fun{X}{ W}{TW}, \hspace{1em} X(p)(\alpha):=\frac{\partial \alpha (p)}{\partial t},
\end{equation}
where  $\alpha\in {\cal W}$, $p\in W$. In other words,  the d-space $\ww$ with
boundary $\partial W$
  is    time oriented with respect  to the cosmological time $t$.

\begin{lem} \label{Ypole} The mapping
$\fun{X}{ W}{TW}$, $X(p)(\alpha):=\partial\alpha (p)/\partial t$ is a vector field
tangent to $(W, {\cal W})$ if and only if  for every $t\in D_a$, $\dot{a}(t)$ is
finite and $\dot{a}(t)\neq 0$. \hfill $\blacksquare$
\end{lem}
\begin{dow}A vector field is tangent to $(W, {\cal W})$ if its value $X(p)(\alpha)$ is finite  for every
 $\alpha\in {\cal W}$ and $p\in W$.
It is enough  to check this property on the generators
$\alpha_1,\alpha_2,...,\alpha_5$ since the d-space $(W, {\cal W})$ is finitely
generated.  Straightforward calculations show that the value of the field
$X(p)(\alpha_2)$ is finite for $p\in W$ iff $\dot{a}(t)$ is finite and $\dot{a}(t)\neq
0$ for $t\in D_a$. Then, the value of $X$ on the remaining generators is always
finite.
\end{dow}

\begin{lem}
If \hspace{0.2em} $V:= \lambda X$,  $\lambda\in {\cal W}$, is a smooth vector field
tangent to the d-space $(W, {\cal W})$ and for every $ p\in W^0$ the value of the
function $ \lambda (p)\neq0 $ then $\lambda (p)=0$ for $p\in
\partial W$.
 In other words, the smooth vector field  $V$ has a critical point at the boundary $\partial W$.
\hfill $\blacksquare$\end{lem}

\begin{dow}
 Every smooth function $\gamma \in {\cal W}$ is a local ${\cal W}$-function on $W$
(see Definition \ref{lokalnecfunkcje}). This means that for every $p\in W$ there is
$U_p\in \tau_{\cal W}$ and $f_p\in C^\infty (\mathbb{R}^5,\mathbb{R})$ such that
$\gamma (q) =f_p(\alpha_1(q),\alpha_2(q),...,\alpha_5(q))$ for $q\in U_p$. Therefore,
the value of $\gamma$ in $p=(0,x,y,z)\in \partial W$, where $x,y,z$ are any,   is a
constant function of $x,y,z$: $\gamma
(0,x,y,z)=f_p(\alpha_1(p),\alpha_2(p),...,\alpha_5(p))=f_p(0,0,0,0)=const$.

Now,
let us suppose that  $\lambda (p)\neq 0$ for $p\in
\partial W$ also. Smoothness conditions for the field $V$ have the form: $\widehat{V}(\alpha_i)\in {\cal
W}, i=1,2,...,5$. In particular, the conditions  $\widehat{V}(\alpha_1)\in {\cal W}$
and $\widehat{V}(\alpha_2)\in {\cal W}$ lead to $\dot{a}\in {\cal W}$ and
$\dot{a}\neq 0$ for $p\in W$.  Then, for example,  the  function $\eta
:=\widehat{V}(\alpha_3)/{\lambda \dot{a}}$, $\eta(p)=x$, is a smooth function
($\eta\in {\cal W}$) since $\lambda\in {\cal W} $ and $\lambda (q)\neq 0$ for $q\in
W$. This is a contradiction since this function is not a constant function on
$\partial W$ and therefore it is not generated by means of $\alpha_i$, $i=1,2,...,5$.
\end{dow}


\section{A smooth evolution with respect to  cosmological time}

\label{smoothevolution}

The vector field $X$ defined by  formula (\ref{poleY}) is not  smooth
on $\ww$. Its value on a smooth function is not necessarily  smooth.
This means, that there are  functions $\alpha\in {\cal W}$ such that
$\widehat{X}(\alpha)\notin {\cal W} $, where the derivation
$\fun{\widehat{X}}{{\cal W}}{\mathbb{R}^{W}}$ is given by the formula
$\widehat{X} (\alpha)(p):=X(p)(\alpha)$. But also it can happen that,
for a class of smooth functions, values of $X$ on functions from this
class are smooth.

 According to the definition of  $X$, its restriction $X^0 \equiv X|_{W^0}$
 is a smooth vector field on  $(W^0, {\cal W}^0)$, and therefore its value
$\dot{\alpha}_1^0:=X^0 (\alpha_1^0)=\dot{a}_0$ on the smooth function
$\alpha_1^0:=\alpha_1|_{W^0} $ is smooth:  $\dot{\alpha}_1^0\in {\cal W}^0$. The
generator $\alpha_1 =a$  is by definition a smooth function on $\ww$. According to the
earlier argumentation, the value of  $X$ on the function $\alpha_1$,
$\dot{\alpha}_1:=X(\alpha_1)=\dot{a}$, is not necessarily  smooth in the Sikorski's
sense.
 But, from the physical point of view, it is natural that  the velocity of
the expansion of the universe  is a smooth function  even at the beginning of  time
$t=0$ since a moment later, it is smooth  ( $\dot{\alpha}_1^0\in {\cal W}^0$ )
 both in the classical sense and  Sikorski's sense. Let us distinguish a class of
cosmological models with such a property.

\begin{defin} \label{gladew}
An evolution of  cosmological model  {\rm (\ref{metryka})} is said to be smooth
 from the beginning of  cosmological time if $\dot{\alpha}_1:=\widehat{X}(\alpha_1)=\dot{a}\in {\cal W}$
 and $\dot{a}(t)\neq 0$ for $t\in D_a$.
 We shall also  say  that the cosmological model is smoothly evolving.
\end{defin}

\begin{tw}\label{poleorientujace}
If the flat FRW model is smoothly evolving, then the d-space $(W, {\cal W})$ with
boundary $\partial W$ of this model is time oriented with respect to the cosmological
time $t$. A smooth vector field $\widehat{V}$ defining time orientability on $\ww$
has the form \hspace{0.5em} $\fun{\widehat{V}}{{\cal W}}{{\cal W}}, \hspace{0.5em}
\widehat{V}:=\alpha_1^2
\partial /\partial t.$
 \hfill $\blacksquare$\end{tw}

\begin{dow} Proof consists of verification  whether the following inclusion is satisfied
 $\widehat{V}(\alpha_i)=\alpha_1^2\frac{\partial \alpha_i}{\partial t}\in {\cal W}, i=1,2,...,5  $.
\end{dow}


\section{The simplest smoothly evolving models}\label{rozwgl}

\begin{lem}\label{generowanieab}
A function  $\fun{\varphi}{W}{\mathbb{R}}$ which depends  on the single coordinate $t$
only
 is smooth  $\varphi\in {\cal W}$ if and only if,  for every  $p\in W$, there is a
neighbourhood   $U_p\in \tau_{\cal W}$ and a function $f\in {\cal E}^{(2)}$ such that
$\varphi|_{U_p}=f(\alpha_1,\xi)|_{U_p}$, where
\begin{equation} \xi:=\alpha_1\alpha_2 -\alpha_3^2 -\alpha_4^2 -\alpha_5^2,
\hspace{1em} \xi(t,x,y,z)=a(t)\int_0^t \frac{d\tau}{\dot{a}(\tau)}.
\end{equation}
 \hfill$\blacksquare$
\end{lem}
\begin{dow}
Let a function  $\varphi$ be a smooth one. The d-structure ${\cal W}$ is finitely
generated. Therefore for every $ p\in W$ there are a neighbourhood $U_p\in\tau_{\cal
W}$ and a function $ f^{(5)}\in {\cal E}^{(5)}$ such that
$\varphi|_{U_p}=f^{(5)}(\alpha_1,\alpha_2,...,\alpha_5 )|_{U_p} .$ Since  $\varphi$
is a constant function with respect to $x, y, z$, it has to be generated by means of
such combinations of the generators admissible in the algebra ${\cal W}$ that a
resulting function depends on $t$ only. The evident form of the generators leads to
conclusion that such combinations have the following form  $\alpha_1$ and
$\xi:=\alpha_1\alpha_2 - \alpha_3^2 -\alpha_4^2 -\alpha_5^2$. Therefore   $f^{(5)}$
has  the following structure $f^{(5)}(\alpha_1,\alpha_2,...,\alpha_5
)=f(\alpha_1,\alpha_1\alpha_2 - \alpha_3^2 -\alpha_4^2 -\alpha_5^2 )=f(\alpha_1,
\xi)$, where $f\in {\cal E}^{(2)}$. The  implication in the opposite direction is obvious.
\end{dow}

According to  Lemma  (\ref{generowanieab}), if $\dot{\alpha}_1\in
{\cal W}$ then, in a  neighbourhood $U_p$ of  $p\in W$, the function
$\dot{\alpha}_1$ is of the form
$\dot{\alpha}_1|_{U_p}=f(\dot{\alpha}_1,\xi)|_{U_p}$. Especially
interesting points in $W$ are those not Hausdorff  separated points
from the boundary $p_\bullet\in\partial W$, $p_\bullet =(0, x, y,
z)$.  For a given
 $f$, from the whole class of all possible
neighbourhoods  $U_{p_\bullet}$ such that
$\dot{\alpha}_1|_{U_{p_\bullet}}=f(\dot{\alpha}_1,\xi)|_{U_{p_\bullet}}$  one can
select the maximal neighbourhood  in the sense of inclusion.
 This means in practice that  the maximal neighbourhood has the form
$U_{p_\bullet}=D_a \times \mathbb{R}^3$, where the domain $D_a$ of the scale factor
is a subset of the  set $\{t\in[0,\infty):0\leqslant a(t)<\infty,
0<\dot{a}(t)<\infty\}$ such that $0\in D_a$. In the neighbourhood $U_{p_\bullet}$ the
smoothness condition for $\dot{\alpha}_1$ (see Definition \ref{gladew}) has the form
\begin{equation}\label{rownaniegladkosci}
\dot{a}(t)=f(a(t), a(t)\int_0^t \frac{d\tau}{\dot{a}(\tau)}),
\end{equation}
where
\begin{equation}\label{waruneknaf}
f(0,0)>0, \hspace{1em} f\in{\cal E}^{(2)}.
\end{equation}

Formula (\ref{rownaniegladkosci}) is an equation for $a(t)$ with an initial condition
$a(0)=0$. The function  $f$, is in a principle,  arbitrary.  The only restriction on
$f$ is  condition (\ref{waruneknaf}) which is a consequence   of Lemmas \ref{Ypole},
\ref{generowanieab} and the physical assumption  that the real universe expands from
the initial singularity. The simplest choice is the following function
$$f(z_1,z_2):=\beta + \gamma_1 z_1 +\gamma_2 z_2, \hspace{1em} \beta>0, \hspace{1em}\gamma_1, \gamma_2 \in \mathbb{R}.$$
Now, the smoothness equation (\ref{rownaniegladkosci}) for $a(t)$ has the form
\begin{equation}\label{rownglad}
\dot{a}(t)=\beta +\gamma_1 a(t) + \gamma_2 a(t)\int_0^t \frac{d\tau}{\dot{a}(\tau)}.
\end{equation}
Solutions of (\ref{rownglad}) have to satisfy  the following conditions
\begin{equation}\label{warunki}
a(0)=0,\hspace{0.5em} \dot{a}(0)=\beta>0, \hspace{0.5em} a(t)>0,\hspace{0.5em}
\dot{a}(t)>0 \mbox{ for } t> 0.
\end{equation}

\begin{stw}\label{lemrozw1}
When $\gamma_2=0$ then solutions of the smoothness equation (\ref{rownglad})
satisfying  conditions (\ref{warunki}) have the form
\begin{equation}\label{rozw1}
a(t)=\beta t ,\hspace{0.5em}t\in[0,\infty),\hspace{0.5em} {\it for}\hspace{0.5em}
\gamma_1=0,
\end{equation}
and
\begin{equation}\label{rozw2}
a(t)=\frac{\beta}{\gamma_1}(e^{\gamma_1 t}-1)
,\hspace{0.5em}t\in[0,\infty),\hspace{0.5em} {\it for}\hspace{0.5em} \gamma_1\neq
0.\end{equation}
 \hfill $\blacksquare$\end{stw}
\begin{dow}
Obvious calculus.
\end{dow}

Solution (\ref{rozw1}) represents the well known  model of the universe which expands
with the constant velocity $\dot{a}=\beta$ and which is a solution of the Friedman's
equations with the following equation of state $p=-\varrho/3$. For  $\gamma_1>0$
solution (\ref{rozw2}) is the universe model which is asymptotically (
$t\rightarrow\infty$) the de-Sitter model. The model  expands from the very beginning
with a positive acceleration.  The parameter $\gamma_1$ can be asymptotically
interpreted as a cosmological constant.
  When
$\gamma_1<0$  cosmological model (\ref{rozw2}) describes  an expanding
universe, and the expansion slows down from the very beginning.
  For great  $t$, the size of universe fixes on the
level     $a(t)\thickapprox
\lim_{t\rightarrow\infty}a(t)=\beta/|\gamma_1|$ and   $\dot{a}$ and
$\ddot{a}$ tend to zero when $t\rightarrow \infty$. Such a universe
asymptotically becomes the Minkowski space-time.

In the case $\gamma_2\neq 0$, let us introduce the following auxiliary symbols
$$
\hspace{0.5em}\gamma_1:=\tilde{\gamma}_1\tilde{\gamma}_2/\sqrt{3}, \hspace{0.5em}
\gamma_2:={\rm sgn}(\gamma_2) \beta\tilde{\gamma}_2^2/3,\hspace{0.5em}
 K:=\sqrt{3}H/\tilde{\gamma}_2, \hspace{0.5em} \tilde{\gamma}_2>0,$$
\begin{equation}\label{tildy}
\tilde{a}(K):=\frac{\tilde{\gamma}_2 \hspace{0.2em}a(K)}{\beta\sqrt{3}},
\hspace{0.5em} \tilde{t}(K):=\tilde{\gamma}_2 \hspace{0.2em}t(K)/\sqrt{3},
\end{equation} where $H(t):=\dot{a}(t)/a(t)$.

\begin{stw}\label{lemrozw3}
If $\gamma_2 >0$ then  solutions of  smoothness equation {\rm(\ref{rownglad})}
have the form
\begin{equation}\label{rozw3}
\tilde{a}(K)=(K-\tilde{\gamma}_1-{\rm arccoth} K)^{-1},\hspace{1em}
\tilde{t}(K)=\int_K^\infty\frac{\tilde{a}(y)ydy}{y^2-1},
\end{equation}
where $K\in(K_f, \infty)$, and $K_f$ is a solution of the following equation
\begin{equation}\label{Kf1}
K_f-{\rm arccoth} K_f=\tilde{\gamma}_1,\hspace{1em} K_f\in (1,\infty).
\end{equation}
\hfill $\blacksquare$\end{stw}
\begin{dow}Solution of an elementary differential equation.
\end{dow}
\begin{stw}\label{przysp1}
If $\gamma_2>0$ then
\begin{itemize}
\item[1.] if $\tilde{\gamma}_1\geqslant 0$,  acceleration  $\ddot{a}(t)>0$ for
$t\in(0,\infty)$,
\item[2.] if  $\tilde{\gamma}_1< 0$,  acceleration $\ddot{a}(t)< 0$ for $t\in
(0,t_*)$, $\ddot{a}(t_*)=0$ and $\ddot{a}(t)>0$ for $t\in (t_*,\infty)$, where
$t_*:=t(K_*)$ and $K_*$ is a solution of the following equation
\begin{equation}\label{kstar1}
\tilde{\gamma}_1+\frac{K_*}{K_*^2-1}+{\rm arccoth} K_*=0,\hspace{1em} K_*\in
(K_f,\infty).
\end{equation}
\end{itemize}
\hfill $\blacksquare$\end{stw}
\begin{dow}
By obvious calculation.
\end{dow}

In the case $\gamma_2>0$,  there are two essentially  different scenarios of a smooth
evolution with respect to the cosmological time $t$:
\begin{itemize}
\item[a)] If $\tilde{\gamma}_1\geqslant 0$ the model accelerates from the very beginning and  expands indefinitely.
\item[b)]In the case $\tilde{\gamma}_1<0$,  initially the expansion slows down,
but at the moment $t_*$ the unlimited and infinitely long accelerated expansion is
initiated (see Figure \ref{figure1}).
\end{itemize}

\begin{itemize}
\item[]\begin{figure}[!h] \unitlength 1cm
\begin{picture}(6,6)(-3,0)
\put(-1,-1){\includegraphics{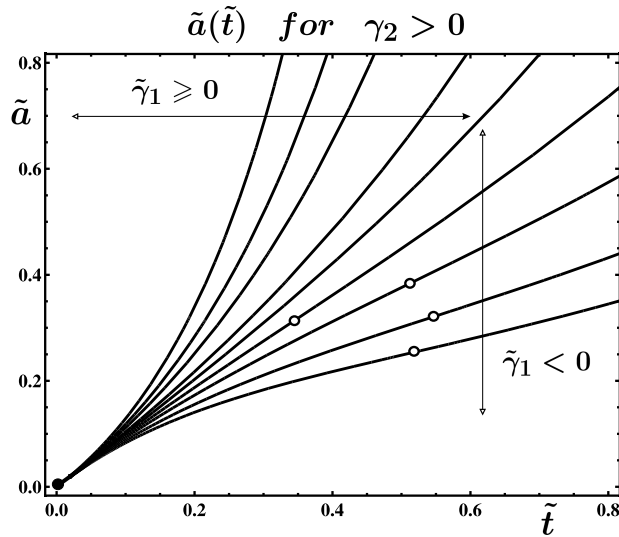}}
\end{picture}
\caption{Scale factor $\tilde{a}(\tilde{t})$ for smooth solutions with $\gamma_2>0$.
When $\tilde{\gamma}_1\geqslant 0$ models  expand with acceleration. For
$\tilde{\gamma}_1<0$ solutions initially expand with a negative acceleration but at a
moment, denoted by a circle on the graph, an accelerated expansion is initiated. The
black point denotes the initial singularity. \label{figure1}}
\end{figure}
\end{itemize}

\begin{stw}\label{lemrozw4}
 If $\gamma_2 <0$, solutions of  smoothness equation {\rm(\ref{rownglad})} have the
 following form
\begin{equation}\label{rozw4}
\tilde{a}(K)=(K-\tilde{\gamma}_1-{\rm arctan} K+\pi/2 )^{-1},\hspace{1em}
\tilde{t}(K)=\int_K^\infty\frac{\tilde{a}(y)ydy}{y^2+1},
\end{equation}
where $K\in (K_f,\infty)$ and $K_f$ is a solution of the following equation
\begin{equation}\label{Kf2}
K_f-{\rm arctan}K_f+\pi/2=\tilde{\gamma}_1,\hspace{0.5em} K_f\in \mathbb{R}.
\end{equation}
\end{stw}
\begin{dow}Solution of an elementary differential equation.
\end{dow}

\begin{stw}\label{oskrz}
If $\gamma_2<0$ and $\tilde{\gamma}_1< \pi/2$, then  cosmological model
 {\rm (\ref{metryka})} has a final curvature singularity at $t_s<\infty$.
 The set $[0,t_s)$ is a domain of the scale factor
$a(t)$, where $t_s:=t(0)$ and the  function $t(K)$ is given by formulae   {\rm
(\ref{rozw4}) and (\ref{tildy})}.
\end{stw}
\begin{dow} Proposition is the result of a fact that some of components of the curvature tensor are undefinite
 at $t_s$, because
$\ddot{a}(t)\rightarrow -\infty $ when $t\rightarrow t_s^-$.
\end{dow}

According to Propositions \ref{lemrozw3} and \ref{lemrozw4}, solutions of smoothness
equation {\rm(\ref{rownglad})} are defined in the domain  $(K_f,\infty)$. But  in the
case of solutions discussed in the Proposition \ref{oskrz}, there appears an
additional restriction for the domains of  $a(K)$ and $t(K)$ coming from the
geometrical interpretation of $a(K)$ as the scale factor for a flat FRW model. The
final curvature  singularity ends the evolution of the model.
\begin{stw}\label{przysp2}
If $\gamma_2<0$ then
\begin{itemize}
\item[1.] if $\tilde{\gamma}_1\geqslant  \pi/2$,  acceleration
$\ddot{a}(t)>0$ for $t\in (0,\infty)$,
\item[2.] if $0<\tilde{\gamma}_1< \pi/2$, the scale factor $a(t)$ has the inflexion point at
 $t_*$ ($\ddot{a}(t_*)=0$), $\ddot{a}(t)>0$ for $t\in [0,t_*)$ and
$\ddot{a}(t)<0$ for $t\in (t_*, t_s)$,
\item[3.] if $\tilde{\gamma}_1\leqslant  0$,  acceleration $\ddot{a}(t)<0$ for $t\in [0,
t_s)$,
\end{itemize}
where  $t_*:=t(K_*)$. The quantity  $K_*$ is a solution of the following equation
\begin{equation}
\frac{K_*}{K_*^2 + 1}-{\rm arctan} K_*+\pi/2 =\tilde{\gamma}_1.
\end{equation}
\hfill $\blacksquare$\end{stw}
\begin{dow} By properties of the function given by formulae (\ref{rozw4}).
\end{dow}

If $\gamma_2<0$, the smooth evolution of the universe with respect to the
cosmological time  $t$ can proceed on three  different  ways:
\begin{itemize}
\item[a)] If $\tilde{\gamma}_1\geqslant  \pi/2$, a smooth accelerated evolution
 starts from the initial singularity. The acceleration goes on
continuously during an infinite period of time.
\item[b)] For  $\tilde{\gamma}_1\in (0,\pi/2)$,   these smoothly evolving
universes initially accelerate but the acceleration is slowing down so as to change, at $t=t_*$,
into a deceleration.  Smooth evolution ends at the  final curvature singularity
within the finite period of time
$[0,t_s)$. These models have two singularities: the
initial and  final one.
\item[c)] Models with   $\tilde{\gamma}_1\leqslant 0$ start their evolution in the Big-Bang and decelerate.
 The rate of expansion is strongly slowing down
and  these models end their evolution at
curvature singularities in a finite time $t_s$ . Models of this  class  have also two singularities.
\end{itemize}

\begin{itemize}
\begin{figure}[!h]
\item[]
\unitlength 1cm
\begin{picture}(7,7)(-3.5,0)

\put(-1,-0.5){\includegraphics{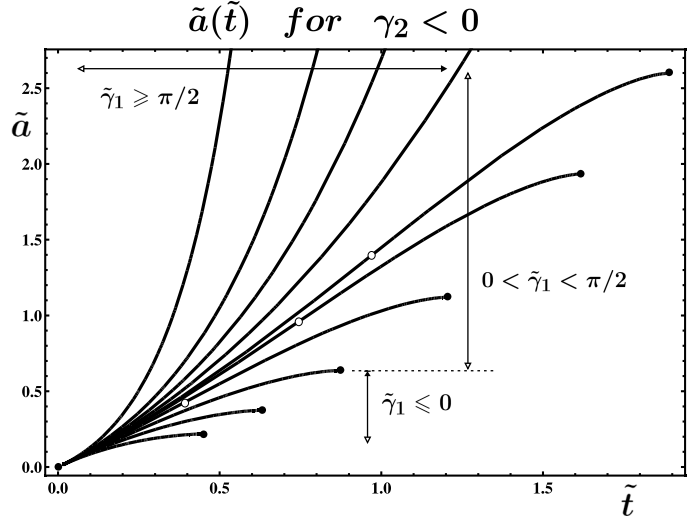}}
\end{picture}
\caption{ Scale factors $\tilde{a}(\tilde{t})$ for smoothly evolving models with
$\gamma_2<0$. Curves on the graph  with $\tilde{\gamma}_1\geqslant \pi/2$ represent
accelerated solutions. For models with $\tilde{\gamma}_1\in (0,\pi/2)$  initially
accelerating expansion is slowing down. At the moment $t_*$ (circles on the graph) the
acceleration changes into a strong deceleration. When $\tilde{\gamma}_1\leqslant0$
solutions expand with  negative acceleration. The black points on the graph denote
initial and final curvature singularities. \label{figure2}}
\end{figure}
\end{itemize}
\pagebreak


\section{Primordial matter and   dark energy }\label{dark}

   Let us assume that solutions
of the smoothness equation represent cosmological models. Then the Friedman equations
\begin{equation}
p=-2\ddot{a}/a-\dot{a}^2/a^2,  \hspace{2em} \varrho=3\dot{a}^2/a^2,
\end{equation}
can serve as a definition of a pressure $\bar{p}$ and  energy density $\bar{\varrho}$
of a kind of cosmological fluid which causes the smooth evolution of models, where
$p=\kappa\bar{p}$ and $\varrho=\kappa\bar{\varrho}$. In the present paper this fluid
is called the cosmological primordial fluid, or  primordial fluid for brevity.

\begin{stw}\label{rownstanu1a}\label{lemrownstanu1}
If $\gamma_1\in\mathbb{R}$ and $\gamma_2=0$ then the equation of state for the
primordial fluid has the form
\begin{equation}\label{rownstanu1}
p=-\frac{1}{3}\rho-\frac{2\gamma_1}{\sqrt{3}}\sqrt{\rho}.
\end{equation}
In addition
\begin{itemize}
\item[a)]the energy density $\varrho$ is a decreasing function of  cosmological time
and $\lim_{t\rightarrow 0^+}\varrho (t)=\infty $,
\item[b)] at the initial moment  $p(0)=\lim_{t\rightarrow 0^+}p(t)=-\infty $, and the remaining details
of the dependence  $p(t)$ are shown in Figure {\rm \ref{figure5}},
\item[c)]
\begin{tabular}{|c|c|c|c|c|}\hline
   $\gamma_2=0$ & $\varrho$ & $p$ & {\rm SEC} & {\rm WEC }\\ \hline\hline
  $\gamma_1>0$ & $\lim_{t\rightarrow\infty}\varrho(t)=3\gamma_1^2$ & $\lim_{t\rightarrow\infty}p(t)=-3\gamma_1^2$ & no & yes
  \\ \hline
  $\gamma_1\leqslant0$ & $\lim_{t\rightarrow\infty}\varrho(t)=0$ & $\lim_{t\rightarrow\infty}p(t)=0$ & yes & yes \\ \hline
\end{tabular}
\end{itemize}
\hfill $\blacksquare$\end{stw}
\begin{dow}
An elementary calculus.
\end{dow}

The abbreviations  SEC and WEC on the above and next tables denote the
strong and  week energy conditions and the statements below are
answers  to the  question of whether the strong or the week energy
conditions are satisfied.

\begin{itemize}
\item[]
\begin{figure}[!h]
\unitlength 1cm
\begin{picture}(6,6)(-3.5,0)
\put(-1,-0.8){\includegraphics{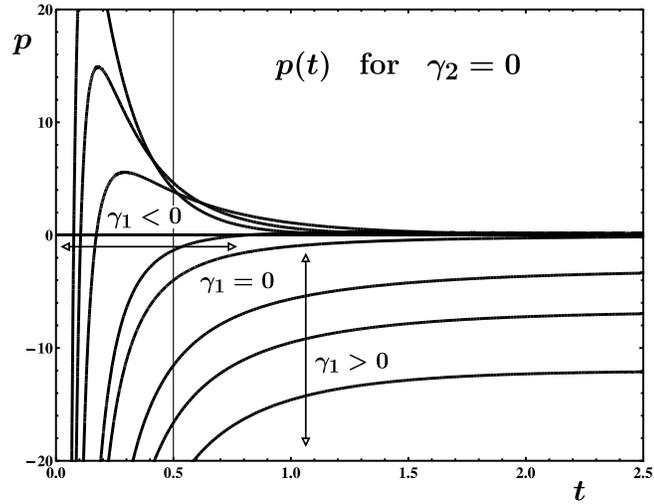}}
\end{picture}
\caption{ Dependence $p(t)$ for $\gamma_2=0$. In the case $\gamma_1<0 $ the pressure
has  a single positive maximum. \label{figure5}}
\end{figure}
\end{itemize}

In the case $\gamma_2\neq 0$ it is convenient to introduce the following  abbreviations
\begin{equation}
\tilde{p}:=p\tilde{\gamma}_2^{-2}, \hspace{2em}
\tilde{\varrho}:=\varrho\tilde{\gamma}_2^{-2}.
\end{equation}
\begin{stw}\label{gama2gt}\label{lemrownstanu2}
If $\gamma_2 > 0$ and $\tilde{\gamma}_1\in \mathbb{R}$ then the equation of state of
the primordial fluid has the form
\begin{equation}
\tilde{p}=-\frac{2}{3}-\frac{1}{3}\tilde{\varrho} -\frac{2}{3}(\tilde{\gamma}_1+{\rm
arccoth}\sqrt{\tilde{\varrho}} \hspace{0.2em})
(\sqrt{\tilde{\varrho}}-1/\sqrt{\tilde{\varrho}}).
\end{equation}
In addition
\begin{itemize}
\item[a)]  energy density $\tilde{\varrho}$ is an increasing function of time and
  \hspace{0.1em}$\lim_{t\rightarrow 0^+}\tilde{\varrho}(t)=+\infty$,
$\lim_{t\rightarrow\infty}\tilde{\varrho}(t)=K_f^2$ where $K_f$ is a solution of
equation {\rm (\ref{Kf1})},
\item[b)]  pressure  at the beginning and  end of the evolution is
$\lim_{t\rightarrow 0^+}\tilde{p}(t)=-\infty$,
$\lim_{t\rightarrow\infty}\tilde{p}(t)=-K_f^2$, and the remaining details of
dependence   $\tilde{p}(t)$ are shown in Figure {\rm \ref{figure3}},
\item[c)] the week energy condition is satisfied during the whole evolution,
\item[d)] for $\tilde{\gamma}_1\geqslant 0$ the strong energy condition is broken down,
\item[e)] if $\tilde{\gamma}_1<0$, the strong energy condition is satisfied for \hspace{0.1em}$t\in
(0,t_*)$. For remaining $t>t_*$ this condition is broken down. The moment $t_*$ is defined
in Proposition {\rm \ref{przysp1}}
\end{itemize}
\end{stw}
\begin{dow}
An elementary calculus.
\end{dow}

\begin{itemize}\item[]
\begin{figure}[!h]
\unitlength 1cm
\begin{picture}(6,6)(-3.5,0)

\put(-1,-1){\includegraphics{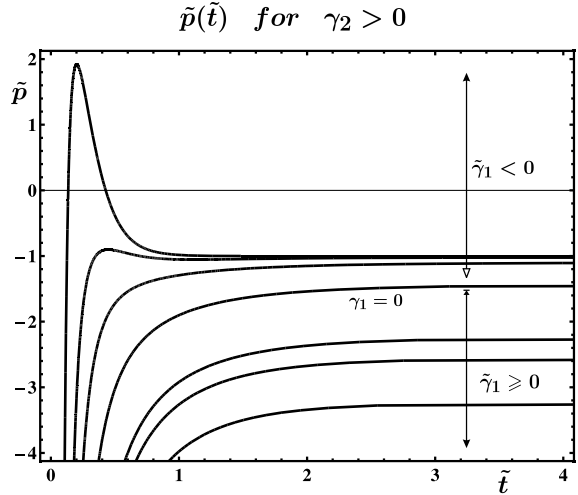}}
\end{picture}
\caption{For $\tilde{\gamma}_1\gtrsim -1.35$,  pressure is a decreasing function
of  cosmological time. For remaining  $\tilde{\gamma}_1$, function
$\tilde{p}(\tilde{t})$ has both the maximum and  minimum. The minimum is not well visible
on the graph.\label{figure3}}
\end{figure}
\end{itemize}
\begin{stw} \label{gama2lt}\label{lemrownstanu3}
If $\gamma_2 < 0$ and $\tilde{\gamma}_1\in \mathbb{R}$ then the equation of state of
the primordial fluid has the following form
\begin{equation}
\tilde{p}=+\frac{2}{3}-\frac{1}{3}\tilde{\varrho} -\frac{2}{3}(\tilde{\gamma}_1-{\rm
arccot}\sqrt{\tilde{\varrho}} \hspace{0.2em})
(\sqrt{\tilde{\varrho}}+1/\sqrt{\tilde{\varrho}}\hspace{0.2em}).
\end{equation}
Additionally
\begin{itemize}
\item[a)]  energy density  $\tilde{\varrho}$ is a decreasing function of the cosmological time,
and $\lim_{t\rightarrow 0^+}\tilde{\varrho}(t)=+\infty$,
\item[b)]  initial pressure is $\tilde{p}(0):=\lim_{t\rightarrow 0+}\tilde{p}(t)=-\infty$,
and the remaining  details of dependence  $\tilde{p}(t)$ are shown in the Figure {\rm
\ref{figure4}},
\item[c)]
\begin{tabular}{|c|c|c|c|c|}\hline
  $\gamma_2 < 0$ & $\tilde{\varrho}$ & $\tilde{p}$ &{\rm SEC} & {\rm WEC }\\ \hline\hline
  $\tilde{\gamma}_1\geqslant \pi/2$ & $\lim_{t\rightarrow \infty}\tilde{\varrho}(t)=K_f^2$ &
      $\lim_{t\rightarrow \infty}\tilde{p}(t)=-K_f^2$ & no & yes \\ \hline
  $0<\tilde{\gamma}_1< \pi/2$ &  $\lim_{t\rightarrow t_s}\tilde{\varrho}(t)=0$ & $\lim_{t\rightarrow t_s}\tilde{p}(t)=+\infty$ & no/yes & yes \\ \hline
  $\tilde{\gamma}_1\leqslant 0$ & $\lim_{t\rightarrow t_s}\tilde{\varrho}(t)=0$ & $\lim_{t\rightarrow t_s}\tilde{p}(t)=+\infty$ & yes & yes \\ \hline
\end{tabular},
\item[d)] if $0<\tilde{\gamma}_1< \pi/2$, the cosmological fluid  violates the strong energy condition
for $t\in (0,t_*)$. In the remaining range $t \in (t_*,t_s)$, the {\rm SEC} is
satisfied.
\end{itemize}
Quantities  $K_f$, $t_s$ and $t_*$ are defined  in Propositions  {\rm
\ref{lemrozw4}},\hspace{0.2em}{\rm \ref{oskrz}} and {\rm \ref{przysp2}}.
\end{stw}
\begin{dow}
Elementary calculations.
\end{dow}

\begin{itemize}\item[]
\begin{figure}[!h]
\unitlength 1cm
\begin{picture}(6,6)(-3.5,0)

\put(-1,-1){\includegraphics{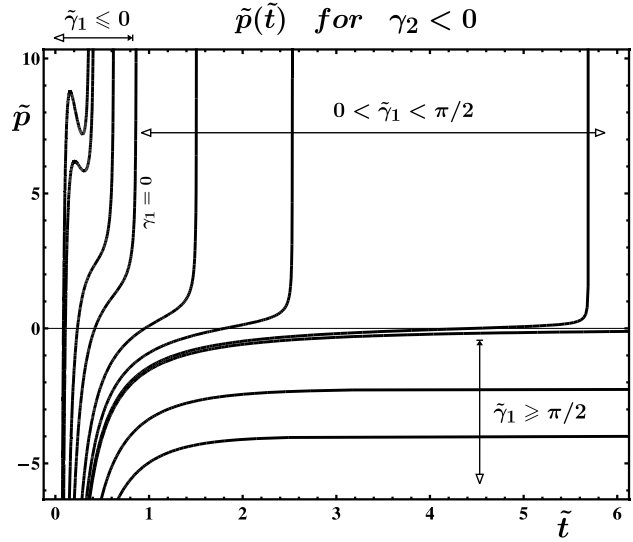}}
\end{picture}
\caption{
The graph shows the great qualitative differences in the $\tilde{p}(\tilde{t})$ dependence
for various ranges of $\tilde{\gamma}_1$.  \label{figure4}}
\end{figure}
\end{itemize}

The simplest solution, $\gamma_1= \gamma_2=0$, of  smoothness equation
(\ref{rownglad}) represents a model filled with the primordial fluid with the
equation of state  $p=-1/3 \varrho$. In the present paper, this fluid is called a
$\gamma_0$-matter. In the case of the following parameters system $\{\gamma_1\neq 0,
\gamma_2=0\}$, the primordial fluid consists of the $\gamma_0$-matter enriched by a
material ingredient connected with the generator
 $\alpha_1(t)=a(t)$  in formula (\ref{rownglad}).
This enriched primordial fluid we  call
 a $\gamma_1$-matter when $\gamma_1<0$, or
a $\gamma_1$-energy when $\gamma_1>0$. Similarly, in the case of the following
parameter system  $\{\gamma_1=0, \gamma_2\neq 0\}$, the primordial fluid composed of
the $\gamma_0$-matter and a  matter connected with the generator $\alpha_2$, through
the function $\xi$  in formula (\ref{rownglad})  , we call a $\gamma_2$-matter when
$\gamma_2<0$, or a $\gamma_2$-energy when $\gamma_2>0$.

Taking into account Proposition \ref{rownstanu1a}   $\gamma_1$-energy
has the properties of a dark energy. This energy causes the expansion
to accelerate.  During the evolution  acceleration grows  to infinity.
After an infinitely long evolution
 pressure and  energy density reach the finite values  $p_f=-3\gamma_1^2$ and
$\rho_f=3\gamma_1^2$ respectively. Let us notice that then the
following equation of state for the cosmological constant is satisfied,
$p_f=-\varrho_f$.  The $\gamma_1$-energy reaches this property at last
stage of the evolution.

In contrast to  $\gamma_1$-energy,  $\gamma_1$-matter satisfies the strong energy
condition during the whole evolution.  It has an attraction property. Therefore the  expansion of
the cosmological model is slowing down in such a manner that at the last stages of the
evolution the model becomes  static. The $\gamma_1$-matter changes its properties during the
evolution.  Initially,  it has a negative pressure. But later it transforms itself
into a kind of matter with a positive pressure.  At the last stages of the evolution the
pressure and the energy density of  $\gamma_1$-matter become zero: $p_f=0$ and
$\varrho_f=0$.  After an infinite period of time since the Big-Bang this cosmological
model becomes, in an asymptotic sense, the Minkowski space-time.

A  model of the universe filled with  $\gamma_2$-energy monotonically
accelerates. Initially, the jostling property of $\gamma_2$-energy has
a small influence on the expansion but its inflationary power is
disclosed at the last stages of evolution of the model. The initially
infinite energy density strongly decreases and at the end of the
evolution is on the level of  $\varrho_f=\tilde{\gamma}_2^2 K_{0f}^2$,
where $K_{0f}\approx 1.19$ is a solution of equation (\ref{Kf1}) for
$\tilde{\gamma}_1=0$. A negative pressure rapidly grows from $-\infty$
to a finite level of $p_f=-\tilde{\gamma}_2^2 K_{0f}^2$. At the end of
the evolution the equation of state is as for  the cosmological
constant:  $p_f=-\varrho_f$. During the whole evolution the strong
energy condition is violated. $\gamma_2$-energy can be interpreted as a
dark energy of different kind then   $\gamma_1$-energy. Details can be
found in Figures \ref{figure1},  \ref{figure3} and in Proposition
\ref{gama2gt}.

 $\gamma_2$-matter has a strong attraction property and therefore the expansion of a
model with such a fluid is rapidly slowing down. Acceleration quickly decreases
from $0$ to $-\infty$ in a finite period. At the and of the evolution the model is stopped
 $\dot{a}(t_f)=0$, and its scale factor reaches the maximal,  finite value. The final
 curvature  singularity ends the evolution.
Properties of  $\gamma_2$-matter are changing during the evolution. Pressure rapidly
increases from $-\infty$ to $+\infty$ in the finite period $[0,t_f)$. Simultaneously,
the energy density decreases from $+\infty$ to zero independently of the fact that
the scale factor is  finite ($a(t_f)<\infty$) at $t_f$. This kind of matter has very
interesting properties at the end of the evolution: it has a slight
energy density but simultaneously  a huge positive pressure. Details can be found in
Figures \ref{figure2},  \ref{figure4} and in Proposition \ref{gama2lt}.

\emph{A mixture of  $\gamma_1$-energy and  $\gamma_2$-energy }($\gamma_1>0 $,
$\gamma_2>0$). During the whole period of evolution of this model  the mixture has
the properties of a dark energy. Both components   interact with each other causing
increased acceleration. In the last stages of the infinitely long evolution, the
equation of state for the mixture has the form of equation of state for the
cosmological constant $\varrho_f=-p_f=\tilde{\gamma}_2^2K_f^2$, where $K_f$ is a
solution of  equation (\ref{Kf1}).

\emph{A mixture of  $\gamma_1$-matter and  $\gamma_2$-matter} ($\gamma_1\leqslant 0
$, $\gamma_2<0$). This kind of the primordial fluid satisfies the strong energy
condition during the whole finite period of evolution. The mixture is a fluid with
interacting components. At the final singularity  $\gamma_2$-matter absorbs, in a
sense, $\gamma_1$-matter and finally the mixture  vanishes, $\varrho_f=0$ at an
infinite pressure. An admixture of $\gamma_1$-matter into $\gamma_2$-matter shortens
the life time of the cosmological model.

 \emph{A mixture of
$\gamma_1$-matter and  $\gamma_2$-energy} ($\gamma_1\leqslant 0 $, $\gamma_2>0$).
Components of the mixture are interacting fluids. The beginning of the model
evolution is dominated by $\gamma_1$-matter. The universe expands with a negative
acceleration and the mixture satisfies the strong energy condition. But the influence
of  $\gamma_2$-energy is still rising.  For $t>t_*$ the evolution is dominated by
$\gamma_2$-energy. Since $t=t_*$, the mixture has properties of a dark energy and
changes  the further evolution  into accelerated expansion. During final stages of the
evolution the equation of state has the following form
$\varrho_f=-p_f=\tilde{\gamma}_2^2K_f^2$, where   $K_f$ is a solution of equation
(\ref{Kf1}).

{The evolutionary behaviour of the model is extremely
 interesting because such an evolution is qualitatively consistent with the observational data of Ia
 type supernovae \cite{RiessAcceleration, PerlAcceleration}.
Preliminary quantitative investigations of the consistency  of the discussed model
have been carried out with the help of the  $\bar{H}_{obs}(z)$ dependence published
in  \cite{SimonVerde}.} Results of the best-fit procedure depend on $\bar{H}_0$. For
$70.6\leqslant\bar{H}_0\leqslant 77.8\hspace{0.2em}{\rm km{\hspace{0.2em} } s^{-1}
Mpc^{-1}}$ \cite{Riess} the best-fit parameters are in the range $\tilde{\gamma}_1\in
{\rm [-1.829, \hspace{0.1em}-1.075]}$, $\tilde{\gamma}_2\in {\rm [2.82
,\hspace{0.1em} \rm 4.011]}$ $\times 10^{-4}{\rm Mpc^{-1}}$ and $\chi_{min}^2\in {\rm
[8.66,\hspace{0.1em} 9.76]}$ (see also Figure \ref{figure6}).  Values of parameters
$\tilde{\gamma}_1$ and $\tilde{\gamma}_2$ enable us to find: an age of universe
$\bar{t}_0\in {\rm [13.561, \hspace{0.1em}14.241]\times 10^{9}\hspace{0.1em}y }$, the
moment of the acceleration beginning  $\bar{t}_*\in {\rm [7.427,
\hspace{0.1em}8.471]\times 10^{9}\hspace{0.1em}y }$, the  Hubble constant in the
acceleration moment $\bar{H}_*\in {\rm [106.31, \hspace{0.1em}108.86]\times
km{\hspace{0.2em} } s^{-1} Mpc^{-1} }$ and redshift $z_*\in {\rm [0.648, 0.743]}$,
where
$$z(K):=\frac{a_0\tilde{\gamma}_2}{\sqrt{3}\beta\tilde{a}(K)}-1,\hspace{0.2em} \bar{t}_0=c^{-1}t(K_0),
\hspace{0.2em} \bar{t}_*=c^{-1}t(K_*),\hspace{0.2em}
K_0:=\frac{\sqrt{3}}{c\tilde{\gamma}_2}\bar{H}_0.$$ Quantities $K$,  $\tilde{a}(K)$,
$t(K)$ are given by formulas (\ref{tildy}) and (\ref{rozw3}), and $K_*$ is a solution
of equation  (\ref{kstar1}). In the present Section  bar over quantities denotes that
we use the systems of units in which $c\neq 1$.

\begin{itemize}\item[]
\begin{figure}[!h]
\unitlength 1cm
\begin{picture}(7,7)(-4,0)

\put(-1,-0.5){\includegraphics{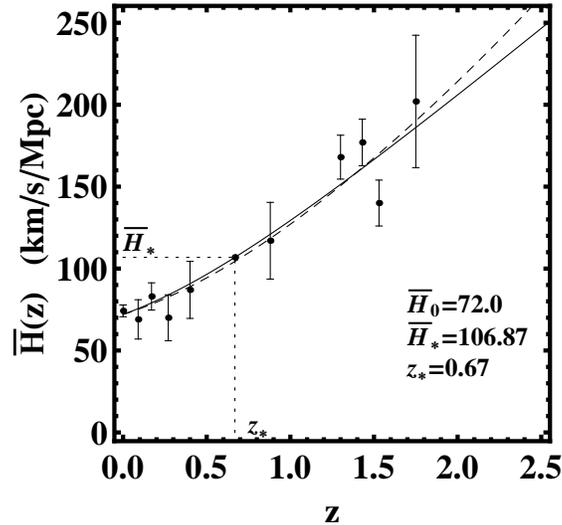}}
\end{picture}
\caption{Comparison of the theoretical $\bar{H}(z)$ dependence for the smoothly
evolved model ($\gamma_1<0$ and $\gamma_2>0$) with $\bar{H}_{obs}(z)$ for
$\bar{H}_0=72.0$ ${\rm km{\hspace{0.2em} } s^{-1} Mpc^{-1}}$. The solid line
represents the smoothly evolved model     with the best-fit parameter values
$\tilde{\gamma}_1={\rm -1.241}$, $\tilde{\gamma}_2={\rm
0.000312\hspace{0.1em}Mpc^{-1}}$ and $\chi^2_{min}={\rm 8.79}$. In this case
$\bar{t}_0={\rm 14.103\times 10^9\hspace{0.1em} y}$,  $\bar{t}_*={\rm 8.241\times
10^9\hspace{0.1em} y}$, $\bar{H}_*=106.87$ ${\rm km{\hspace{0.2em} } s^{-1}
Mpc^{-1}}$ and $z_*=0.668$. The dashed line represents the best-fit of the
$\Lambda$CDM model.\label{figure6}}
\end{figure}
\end{itemize}
\emph{A mixture of  $\gamma_1$-energy and  $\gamma_2$-matter} ($\gamma_1> 0 $,
$\gamma_2<0$). For $t\in [0,t_*)$ the mixture has properties of  a dark energy and
therefore this model accelerates from the very beginning. But later,
$\gamma_2$-matter component begins to play a bigger and bigger  role. The further
evolution depends on the value of the parameter   $\gamma_1$.

When $\tilde{\gamma}_1\in (0,\pi/2)$, the repulsive properties of $\gamma_1$-energy
are not able to dominate the  evolution and the acceleration is gradually stopped
because of a greater and greater   attractive influence of $\gamma_2$-matter.  The
moment $t_*$, defined in the Proposition  \ref{przysp2},  is the end of acceleration.
Starting from this moment, the expansion is strongly slowing down
 till the final singularity. The behaviour of  pressure is interesting,
 Figure \ref{figure4}, for $t>t_*$. After $t_*$
the mixture behaves as  a dust   ($\tilde{p}\thickapprox 0$).  But later, pressure
rapidly increases  to infinity at the final singularity.   Simultaneously, because of
expansion,  the energy density decreases to zero.

When $\tilde{\gamma}_1\geqslant \pi/2$,  repulsive influence of $\gamma_1$-energy is
dominating during the whole period the infinitely long evolution.  The mixture acts
like
 a dark energy  causing acceleration, independently of the $\gamma_2$-matter presence.
  As for previously considered accelerating models,  the equation of state,
  in the last stages of the evolution, is as the one
    for the cosmological constant, i.e.,
$\varrho_f=-p_f=\tilde{\gamma}_2^2K_f^2$, where $K_f$ is a solution of  equation
(\ref{Kf2}).


\section{Smoothly evolved models in a neighbourhood of singularity}
\label{poblize}

The  function $f$ appearing in the smoothness equation (\ref{rownaniegladkosci}) can
be  expanded into a series in a neighbourhood of the point $(0,0)$: $f(z_1,z_2)=\beta
+ {\partial}_1 f(0,0)\cdot z_1 + {\partial}_2  f(0,0)\cdot z_2 +...$. Then the
smoothness equation
 assumes the following form
$$\dot{a}(t)=\beta + {\partial}_1
f(0,0)\cdot a(t) + {\partial}_2  f(0,0)\cdot a(t)\int_0^t \frac{d\tau}{\dot{a}(\tau)}
+...\hspace{0.2em}.$$ For sufficiently small $t$, or equivalently in a small
neighbourhood of the initial singularity, one can omit higher powers of the expansion
and consider the smoothness equation in  the linear approximation. If one assumes
that $\gamma_1:={\partial}_1 f(0,0)$ and $\gamma_2:={\partial}_2 f(0,0)$, the above
equation, in the linear approximation, is identical with the smoothness equation
(\ref{rownglad}) for the  simplest smoothly evolved models (Section \ref{rozwgl}).

The above observation leads to the conclusion that properties  of the solutions
studied in Sections \ref{rozwgl} and \ref{dark}, for small $t$, are typical for every
smoothly evolving flat FRW model in a small neighbourhood of the initial singularity.
In particular, every smoothly evolving model  during the initial stages of its
evolution is filled with a cosmological fluid which is $\gamma_i$-matter or
$\gamma_i$-energy, $i=1,2$, or one of mixtures described in Section \ref{dark}. These
fields satisfy the following approximate equation of state
$$p=-\frac{\varrho}{3}-\frac{2\gamma_1}{\sqrt{3}}\sqrt{\varrho}-\varepsilon\frac{4}{3}
\tilde{\gamma}_2^2
+\varepsilon\frac{2\gamma_1\tilde{\gamma}_2^2}{\sqrt{3}}\frac{1}{\sqrt{\varrho}}
+\varepsilon^2\frac{4}{9}\frac{\tilde{\gamma}_2^4}{\varrho}+\varepsilon^3\frac{4}{45}\frac{\tilde{\gamma}_2^6}{\varrho^2}+...
,$$ where $\gamma_1\in \mathbb{R}$, $\tilde{\gamma}_2\geqslant 0$, $\varepsilon =
{\rm sign}(\gamma_2)$.


\section{Summary and Discussion } \label{summary}

 a) The discussion of the equivalence principle and its breaking down has led
to the formulation of the d-boundary notion which is, roughly speaking, an
''optimal'' set of points at which the equivalence principle is broken. A space-time
d-manifold with the attached d-boundary is not only a topological space but also an
object with a rich geometrical structure called differential space (see Appendix
\ref{appendix}). The effective construction  of both d-boundary and d-space with
d-boundary for any space-time d-manifold has also been described (see Section
\ref{dsing}).

b) For every flat FRW cosmological model with the initial singularity the d-space
with d-boundary $(\bar{W},\bar{\cal W})$  has been constructed. In this case the
d-boundary is a single point (see Section \ref{dsp}).

c) In the d-space formalism it is possible to extend the concept of time orientability
to the d-boundary of the FRW models. In this way, the intuitive understanding of the
beginning of the cosmological time obtains a precise mathematical form  (see Section
\ref{wekt}). However, not every flat FRW model with the initial singularity has the
well defined beginning of the cosmological time.

 d) In the whole class of flat FRW models with the well defined
beginning of  the cosmological time we distinguish the large subclass of the so-called
smoothly evolved models (see Definition \ref{gladew}). The condition defining this
subclass is called the smoothness equation.  The most practical form of this equation
is given by formula (\ref{rownaniegladkosci}).

e)  The simplest flat FRW models with the well defined beginning of the cosmological
time have been found in the explicit form. This set of models can be divided into six
qualitatively different classes (Lemmas \ref{lemrozw1}, \ref{lemrozw3} and
\ref{lemrozw4}). The most important classes are
\begin{itemize}
\item[$\bullet$] solutions with parameters $\tilde{\gamma}_1<\pi/2$ and  ${\gamma_2<0}$
which have two curvature singularities: the initial singularity and the final
singularity (see  Figure \ref{figure2}),
\item[$\bullet\bullet$] the subfamily with $\tilde{\gamma}_1<0$ and
${\gamma_2>0}$, being qualitatively consistent with the observational
$\bar{H}_{obs}(z)$ data from \cite{SimonVerde}. The quantitative consistency with the
data is discussed in Section \ref{dark} and is shown in Figure \ref{figure6}. The
level of the consistency is similar to that for the $\Lambda$CDM model.
\end{itemize}
f) The Friedman equation without cosmological constant, in application to the
simplest solutions, may serve as a definition of energy density $\bar{\varrho}$ and
pressure $\bar{p}$ of a cosmological fluid (primordial fluid) which causes the smooth
evolution. This strategy enables us to find  main phenomenological properties (in
particular, the equation of state) of this primordial fluid (see Lemmas
\ref{lemrownstanu1}, \ref{lemrownstanu2} i \ref{lemrownstanu3}).

g) In Section \ref{dark} we present an interpretation of the primordial fluid as a
mixture of more elementary interacting primordial fluids $\gamma_i$-matter and
$\gamma_i$-energy, $i=1,2$.

h) From the  analysis  of smoothness equation (\ref{rownaniegladkosci}) in a
neighbourhood of the initial singularity, i.e for small $t$, one can conclude (see
Section \ref{poblize}) that in  earlier stages of the evolution every smoothly
evolving flat FRW model is filled with a primordial cosmological fluid with
properties characteristic for particular solutions found in this paper (Sections
\ref{rozwgl} and \ref{dark}).

i) It is very surprising  that without any assumption concerning physical nature of
the cosmological  fluid , among the simplest solutions of the smoothness equation
(\ref{rownaniegladkosci}), it is possible to find  models consistent with the
observed evolution of the Universe  (see Figure \ref{figure6} ). The existence of the
well defined beginning of the cosmological time was the only requirement which has
been assumed. Primordial mixture of fluids $\gamma_1$-matter and $\gamma_2$-energy is
a consequence of this simple assumption. However, we cannot expect that every detail
of the cosmic evolution can be determined in this way. The material ingredients such
as radiation, dust and dark matter should also be taken into account.

\dzieki This publication was made possible through the support of a
grant from the John Templeton Foundation.


\appendix
\section{Sikorski's differential spaces}
\label{appendix}


Let ${\cal C}$ be a non empty family of real functions defined on a set  $M$. The
family ${\cal C}$ generates on  $M$ a topology denoted by the symbol $\tau_{\cal C}$.
It is the weakest topology on $M$ in which every function from ${\cal C}$ is
continuous.

Let $A\subset M$ be a subset of $M$ and  let the symbol ${\cal C}|A$ denotes the set
of all functions belonging to  ${\cal C}$ restricted to $A$. On $A$ one can define the
 induced topology $\tau_{\cal C}\cap A=\tau_{{\cal C}|A}$. The topological space
 $(A,\tau_{{\cal C}|A})$ is a topological subspace of $(M,\tau_{\cal C})$.

Next, we introduce two key notions in the d-spaces theory: a) the closure of  ${\cal
C}$ with respect to localization and b) the closure of  ${\cal C}$ with respect to
superposition with smooth functions from ${\cal E}^{(m)}:={\cal C}^\infty
(\mathbb{R}^m,\mathbb{R})$, $m=0,1,2,...$.

\begin{defin} \label{lokalnecfunkcje}
A function $\fun{\gamma}{A}{\mathbb{R}}$  is said to be a local \hspace{0.1em} ${\cal
C}$-function on a subset $A\subset M$ if, for every $ p\in A$, there is a
neighbourhood $U_p\in \tau_{{\cal C} |A}$ and a function $\phi\in {\cal C}$ such that
$\gamma|{U_p}=\phi|{U_p} $. The set of all local  ${\cal C}$-functions on $A$ is
denoted by ${\cal C}_A$.
\end{defin}

It is easily to check that in general ${\cal C}|_A\subset {\cal C}_A$. In particular
${\cal C}\subset {\cal C}_M$.

\begin{defin} A family ${\cal C}$ of real functions   on a set $M$
is said to be closed with respect to localization if \hspace{0.1em} ${\cal C}= {\cal
C}_M$.
\end{defin}

\begin{defin} A family of functions ${\cal C}$ is closed with respect to superposition
with smooth functions from  ${\cal E}^{(m)}$, $m=0,1,2,...$, if for every function
$\omega\in {\cal E}^{(m)}$ and for every system of $m$ functions $\varphi_1,
\varphi_2,..,\varphi_m\in {\cal C}$, the superposition
$\omega(\varphi_1,\varphi_2,...,\varphi_m)$ is a function from ${\cal
C}$;\hspace{0.1em} $\omega(\varphi_1,\varphi_2,...,\varphi_m)\in {\cal C}$.
\end{defin}
The above described system of concepts make possible to define an object (a d-space)
which is a commutative generalisation of the d-manifold concept.

\begin{defin}\label{defdsp}
A pair   $(M, {\cal C})$, where $M$ is a set of points and ${\cal C}$ a family of
real functions on $M$, is said to be a differential space (d-space for brevity) if
\begin{itemize}
\item[1.] ${\cal C}$ is closed with respect to localization, ${\cal C}= {\cal C}_M$,
\item[2.] ${\cal C}$ is closed with respect to superposition with smooth functions from ${\cal
E}^{(m)}$.
\end{itemize}
The family ${\cal C}$ is called a differential structure on  $M$ (d-structure for
brevity) and the set  $M$  a support of the d-structure ${\cal C}$. Functions $\varphi
\in {\cal C}$ are called  smooth functions.
\end{defin}

Every d-space is simultaneously a topological space with the topology $\tau_{\cal C}$
given by the d-structure ${\cal C}$ in the standard way. The d-structure itself, with
the usual multiplication, is a commutative algebra. The notion of smoothness, given by
the condition $\varphi \in {\cal C}$, is an abstract generalization of the smoothness
notion for functions defined on  $\mathbb{R}^n$. Differential structure, by
definition, is a set of all  smooth functions on $M$. There are no other smooth
functions on $M$. This class of smooth  functions may consists of functions which are
not smooth in the traditional sense. This  is a great advantage of the d-spaces
theory. The simplest example of a d-space is the n-dimensional Euclidean d-space
$(\mathbb{R}^n, {\cal E}^{(n)})$, where ${\cal E}^{(n)}={\cal C}^\infty
(\mathbb{R}^n,\mathbb{R})$.

 There exists a procedure to  construct a d-structure
with the help of a chosen set of real functions on $M$. Let us denote it as ${\cal
C}_0$. The method consists in adding  to a given ${\cal C}_0$ missing functions so as
to satisfy the axioms of the closure with respect superposition with smooth functions
and the closure with respect to localization. The closure with respect to
superposition with  smooth real  functions on $\mathbb{R}^n$, is denoted by
mathematicians  by ${\rm sc}({\cal C}_0)$  and the closure with respect to
localization is denoted by $({\cal C}_0)_M$ (see Definition \ref{lokalnecfunkcje}) or
${\rm lc}({\cal C }_0)$.

It is easy to check that
\begin{lem}\label{ogen} Let ${\cal C}_0$
be a set of real functions defined on a set $M$.  The family of functions ${\cal
C}:={\rm lc}({\rm sc}({\cal C}_0))=({\rm sc}({\cal C}_0))_M$ is the smallest, in the
sense of inclusion,  d-structure on  $M$ containing ${\cal C}_0$.
\end{lem}
 Sometimes one uses the following
abbreviation:  ${\cal C}={\rm Gen}({\cal C}_0):={\rm lc}({\rm sc}({\cal C}_0))=({\rm
sc}({\cal C}_0))_M$.

\begin{defin} The set ${\cal C}_0$ in  Lemma \ref{ogen} is said to be a set of generators.
Functions $\varphi\in{\cal C}_0$ are called generators of the d-structure
\hspace{0.1em}${\cal C}:={\rm lc}({\rm sc}({\cal C}_0))$. If ${\cal C}_0$ is finite
then the d-structure ${\cal C}$ is called finitely generated.
\end{defin}

The method of constructing a d-structure with the help a set of generators is the
great advantage of the Sikorski's theory, especially   in the case of finitely
generated d-spaces such as, for example,  the d-space with singularity associated with
the flat FRW world model.

\begin{defin}
If $(M, {\cal C})$ is a d-space and $A\subset M$ then the d-space $(A, {\cal C}_A)$
is said to be a differential subspace (d-subspace) of the d-space $(M, {\cal C})$.
\end{defin}

The above definition enables  us to determine a d-structure for any subset $A$ of $M$.
It is enough to ''localize''  every function from the d-structure ${\cal C}$ to $A$.
In the case of a finitely generated d-spaces  $(M, {\cal C})$, a simpler situation
occurs. Then ${\cal C}={\rm Gen}({\cal C }_0)$, ${\cal C
}_0:=\{\beta_1,\beta_2,...,\beta_n \}$ , $n\in\mathbb{N}$, where
$\beta_1,\beta_2,...,\beta_n$ are given functions. The d-structure ${\cal C}_A$ is
given in terms of  generators   $\tilde {{\cal C}}_0={\cal C }_0|A$ which is a set of
restrictions of the set ${\cal C }_0$ to $A$. Then ${\cal C}_A={\rm Gen}(\tilde {{\cal
C}}_0)$.

\begin{defin}Let $(M,{\cal C})$ i $(N,{\cal D})$ be a d-spaces.
\begin{itemize}
\item[1.] A mapping $\fun{f}{M}{N}$ is said to be smooth if $\forall \beta\in {\cal D}:
\hspace{0.5em}\beta\circ f\in{\cal C}.$
\item[2.] A mapping $\fun{f}{M}{N}$ is said to be a diffeomorphism from a d-space
$(M,{\cal C})$ to a d-space $(N,{\cal D})$, if it is a bijection from
$M$ to $N$ and both
 mappings
 $\fun{f}{M}{N}$ and $\fun{f^{-1}}{N}{M}$ are smooth. \\In this case,
we say that $(M,{\cal C})$ and $(N,{\cal D})$ are diffeomorphic.
\end{itemize}
\end{defin}

 A smooth mapping $f$ transforms smooth functions on $N$ onto smooth functions on $M$.
The notion of d-spaces diffeomorphism is the key notion from the point of view of the
present paper.  If there is a diffeomorphism $f$ between d-spaces $(M,{\cal C})$ and
$(N,{\cal D})$ then these d-spaces, from the viewpoint the d-spaces theory, are
equivalent.

\begin{defin}\label{locdyf} Let $(M,{\cal C})$ i $(N,{\cal D})$ be  d-spaces.
A d-space $(M,{\cal C})$ is said to be locally diffeomorphic to the d-space $(N,{\cal
D})$ if for every  $p\in M$ there is $U_p\in \tau _{\cal C}$ and a mapping
$\fun{f_p}{U_p}{f_p(U_p)\in \tau_{\cal D}}$ such that $f_p$ is a diffeomorphism
between the d-subspaces $(U_p, {\cal C}_{U_p})$ and  $(f_p(U_p), {\cal D}_{f_p(U_p)})$
of the d-spaces $(M,{\cal C})$ i $(N,{\cal D})$, respectively.
\end{defin}

\begin{defin}\label{dman}
A d-space $(M,{\cal C})$ is said to be an n-dimensional d-manifold, if it is locally
diffeomorphic to the d-space   $(\mathbb{R}^n, {\cal E}^{(n)})$.
\end{defin}

Applying  Definition \ref{locdyf} to Definition \ref{dman}, leads to
condition ($\star$) in Definition \ref{dmanifold}. Local
diffeomorphisms $f_p$ are obviously maps. A set of maps forms an
atlas. It turns out that Definition \ref{dman} is equivalent to the
classical definition of d-manifold  \cite{3bo}.

\begin{tw}\label{image} Let $(M,{\cal C})$ be a finitelly generated d-space with the structure
  ${\cal C}$ generated by a finite set of functions:
  ${\cal C}_0:=\{\beta_1,\beta_2,...,\beta_n \}$, ${\cal C}={\rm Gen}({\cal C}_0)$.
Then the mapping $\fun{F}{M}{\mathbb{R}^n}$,
$F(p):=(\beta_1(p),\beta_2(p),...,\beta_n(p))$ is a diffeomorphism of the d-space
 $(M,{\cal C})$ onto the  d-subspace  $(F(M),{\cal
E}^{(n)}_{F(M)})$ of the d-space $(\mathbb{R}^n, {\cal E}^{(n)})$.
\end{tw}
Proof, see \cite{32af}.

The d-subspace $(F(M),{\cal E}^{(n)}_{F(M)})$ of $(\mathbb{R}^n, {\cal
E}^{(n)})$ is an image of the d-space $(M,{\cal C})$ in the mapping
$F$. Theorem \ref{image} is called the theorem on a diffeomorphism
onto the image and it is important for the construction of the d-space
for the flat FRW model with the initial singularity.

\begin{defin}
A mapping \hspace{0.2em} $\fun{v}{\hspace{0.1em}\cal C}{\mathbb{R}}$ is said to be a
tangent vector to a d-space  $(M,\cal{C})$ at a point $p\in M$ if
\begin{itemize}
\item[1.] $\forall \alpha,\beta \in {\cal C}\forall a,b \in\mathbb{R}:\hspace{0.5em}
v(a\alpha+b\beta)=av(\alpha)+bv(\beta), $
\item[2.]
$\forall\alpha,\beta\in {\cal C}:
\hspace{0.5em}v(\alpha\beta)=v(\alpha)\beta(p)+\alpha(p)v(\beta),\hspace{0.5em} p\in
M.$
\end{itemize}
The set of all tangent vectors to $(M,{\cal C})$ at $p\in M$ is said to be a tangent
vector space to $(M,{\cal C})$ at $p\in M$ and is denoted by $T_pM$. The symbol
\hspace{0.1em}$TM$ denotes the following disjoint sum:
$$TM:=\bigcup_{p\in M}T_pM.$$

\end{defin}

\begin{defin}\label{polewektorowe}
Let  $(M,{\cal C})$ be a d-space. The mapping  $\fun{X}{M}{TM}$ such that $\forall
p\in M, \hspace{0.5em}X(p)\in T_pM$ is said to be a vector field  tangent to $(M,{\cal
C})$.
\end{defin}
With help of a vector field  $\fun{X}{M}{TM}$ one can define the following mapping
$$\fun{\widehat{X}}{{\cal C }}{\mathbb{R}^M}, \hspace{1em}\widehat{X}(\alpha)(p):=X(p)(\alpha),$$
where $\alpha\in{\cal C}$ i $p\in M$. The mapping is  linear  and satisfies the
Leibnitz rule  $\forall \alpha,\beta\in {\cal C}:
\widehat{X}(\alpha\beta)=\widehat{X}(\alpha)\beta+\alpha\widehat{X}(\beta)$.
Therefore,  it is a derivation  and  a global alternative  for the definition of a
vector field (\ref{polewektorowe}).
\begin{defin}\label{polewektorowegladkie}
A vector field  tangent to $(M,{\cal C})$ is said to be smooth if the mapping
$\widehat{X}$ satisfies the condition: $\widehat{X}({\cal C})\subset {\cal C}.$
\end{defin}
\begin{defin}
Let $\fun{f}{M}{N}$ be a smooth mapping. The mapping $\fun{f_{*p}}{T_pM}{T_{f(p)}N}$,
given by the formula $$\forall v\in T_pM, \beta\in {\cal D}:
\hspace{1em}f_{*p}(v)(\beta):=v(\beta\circ f),$$ is said  to be  differential of the
mapping $f$ at the point $p\in M$.
\end{defin}

Let us define the following mapping $\fun{id_A}{A}{M}, \hspace{0.2em} id_A(p)=p $,
where $A\subset M$.

\begin{defin}
A vector field $\fun{Y}{M}{TM}$ on $(M, {\cal C})$ is said to be  tangent to a
d-subspace  $(A, {\cal C}_A)$, if  there is a vector field $\fun{X}{A}{TA}$ on $(A,
{\cal C}_A)$ such that
 $$\forall p\in A: Y(p)=id_{A*p}(X).$$
\end{defin}




\begin{thebibliography}{10}

\bibitem{RiessAcceleration}
Riess A.G, Filippenko A.V., Challis P., Clocchiatti A., Diercks A.,
Garnavich P.M., Gilliland R.L, Hogan C.J, Jha S., Kirshner R.P,
Leibundgut B., Phillips M.M, Reiss D, Schmidt8 B.P.,9, Schommer R.A.,
Smith R.C., Spyromilio J., Stubbs C., Suntzeff N.B and Tonry J.
\newblock {\em Astronomical Journal}, 116:1009, 1998.

\bibitem{PerlAcceleration}
Perlmutter S., Aldering1 G., Goldhaber G.,  Knop R. A., Nugent P.,
Castro P.G, Deustua S., Fabbro S., Goobar A.,  Groom D. E., Hook I. M.,
  Kim A. G.,  Kim M. Y.,  Lee J. C.,  Nunes N. J.,  Pain R.,
 Pennypacker C. R.,  Quimby R.,  Lidman C.,  Ellis R. S.,
Irwin M.,  McMahon R. G.,  Ruiz-Lapuente P.,  Walton N.,  Schaefer B.,
 Boyle B. J.,  Filippenko A. V.,  Matheson T.,  Fruchter A. S.,
Panagia N.,  Newberg H. J. M.,  Couch W. J. and The Supernova Cosmology
Project.
\newblock {\em Astrophysical Journal}, 517:565, 1999.

\bibitem{9bo}
Beem J. and Ehrlich P.
\newblock {\em Global Lorentzian Geometry}.
\newblock Marcel Dekker. Inc. New York and Basel, 1981.

\bibitem{15bo}
Weinberg S.
\newblock {\em Gravitation and Cosmology}.
\newblock Wiley, New York, 1972.

\bibitem{16bo}
Torretti R.
\newblock {\em Relativity and Geometry}.
\newblock Pergamon Press, Oxford, 1983.

\bibitem{17bo}
Heller M. and Raine D.J.
\newblock {\em The Science of Space-Time}.
\newblock Pachart, Tucson, 1981.

\bibitem{kobayashi}
Kobayashi S. and Nomizu K.
\newblock {\em Foundations of Differential Geometry}.
\newblock Interscience Publishers, New York, London, 1962.

\bibitem{aronszajn1}
Aronszajn N.
\newblock Subcartesian and subriemannian spaces.
\newblock {\em Notices of the American Mathematical Society}, 14:111, 1967.

\bibitem{27am}
Marshall Ch.D.
\newblock Calculus on subcartesian spaces.
\newblock {\em Journal of Differential Geometry}, 10:551--574, 1975.

\bibitem{28am}
Mostov M.A.
\newblock The differentiable space structures of Milnor classifying spaces-
  simplicial complex and geometric relations.
\newblock {\em Journal of Differential Geometry}, 14:255--293, 1979.

\bibitem{61am}
Chen~K. T.
\newblock {\em Bulletin of the American Mathematical Society}, 83:831, 1977.

\bibitem{3bo}
Sikorski R.
\newblock {\em Introduction to Differential Geometry}.
\newblock Polish Scientific Publishers, 1972.
\newblock (in Polish).

\bibitem{34af}
R.~Sikorski.
\newblock  Abstract covariant derivative.
\newblock \emph{Colloquium Mathematicum}, 18: 251--272, 1967.

\bibitem{35af}
R.~Sikorski.
\newblock Differential modules.
\newblock \emph{Colloquium Mathematicum}, 24: 45--79, 1971.


\bibitem{12af}
Gruszczak J., Heller M., and Multarzynski P.
\newblock A generalization of manifolds as space-time models.
\newblock {\em Journal of Mathematical Physics}, 29:2576--2580, 1988.

\bibitem{31am}
Sasin W.
\newblock Differential spaces and singularities in differential space-times.
\newblock {\em Demonstratio Mathematica}, 24:601--634, 1991.

\bibitem{rosinger2}
Rosinger E.E.
\newblock Differential algebras with dense singularities on manifolds.
\newblock {\em Acta Applicandae Mathematicae}, 95(3):233--256, 2007.

\bibitem{megied1}
Abdel-Megied M. and Gad R.M.
\newblock On the singularities of reissner-nordstrom space-time.
\newblock {\em Chaos Solitons and Fractals}, 23(1):313--320, 2005.

\bibitem{buchner1}
Buchner K.
\newblock Differential spaces and singularities of space-time.
\newblock {\em General Mathematics}, 5(1-4):53--66, 1995.

\bibitem{heller1}
Heller M. and Sasin W.
\newblock Origin of classical singularities.
\newblock {\em General Relativity and Gravitation}, 31:555--570, 1999.

\bibitem{3af}
Clarke C.J.S.
\newblock On the global ismetric embeding of pseudo-riemannian manifolds.
\newblock {\em Proceedings of the Royal Society of London-A}, 314:417--428,
  1970.

\bibitem{69af}
Gruszczak J.
\newblock Discrete spectrum of the deficit angle and the differential structure
  of a cosmic string.
\newblock {\em International Journal of Theoretical Physics}, 47:2911--2923,
  2008.

\bibitem{30am}
Sasin W.
\newblock On equivalence relations on a differential space.
\newblock {\em Commentationes Mathematicae Universitatis Carolinae},
  29:529--539, 1988.

\bibitem{58am}
Waliszewski W.
\newblock Regular and coregular mappings of differential spaces.
\newblock {\em Annales Polonici Mathematici}, 30:263--281, 1975.

\bibitem{32af}
Sasin W. and Zekanowski Z.
\newblock On locally finitely generated differential spaces.
\newblock {\em Demonstratio Mathematica}, 20:477--486, 1987.

\bibitem{70af}
Geroch R.
\newblock Space-time structure from a global viewpoint.
\newblock In {\em General Relativity and Cosmology, Proceedings of the
  International School of Physics Enrico Fermi}, pages 71--103, 1971.

\bibitem{SimonVerde}
Jimenez~R. Simon~J., Verde~L.
\newblock Constraints on the redshift dependence of the dark energy potential.
\newblock {\em Physical Reviev D}, 71:123001, 2005.

\bibitem{Riess}
Riess A.G., Macri L., Casertano S., Sosey M., Lampeitl H., Ferguson
H.C., Filippenko A.V., Jha S.W., Li W., Chornock R. and Sarkar D.
\newblock {\em Astrophysics Journal}, 699:539, 2009.

\end{thebibliography}

\end{document}